\PassOptionsToPackage{table}{xcolor}
\documentclass[acmsmall,screen,review=false]{acmart}
\usepackage[ruled,vlined,linesnumbered]{algorithm2e}
\usepackage{graphicx} 
\usepackage{adjustbox} 
\usepackage{enumitem}
\usepackage{booktabs}
\usepackage{tabularx}
\usepackage{makecell}
\usepackage{ragged2e}
\usepackage{multirow}
\usepackage{multicol}
\usepackage{wrapfig}  
\usepackage{lipsum}   

\newcommand{\addedchw}[1]{#1}

\newcommand{\addedfinal}[1]{#1}

\usepackage{listings}
\usepackage{graphicx}
\usepackage{adjustbox}
\usepackage{enumitem}
\usepackage{booktabs}
\usepackage{tabularx}
\usepackage{makecell}
\usepackage{multirow}
\usepackage{multicol}
\usepackage{wrapfig}
\usepackage{lipsum}
\usepackage{longtable}
\usepackage{float} 
\usepackage{listings}
\usepackage{booktabs}
\usepackage{subcaption}
\usepackage{colortbl}

\definecolor{tablegray}{gray}{0.95}
\usepackage{tcolorbox}
\lstdefinestyle{promptstyle}{
  basicstyle=\ttfamily\small,
  breaklines=true,
  breakatwhitespace=false,
  postbreak=\mbox{\textcolor{gray}{$\hookrightarrow$}\space},
  xleftmargin=2mm, frame=single,
  framesep=2mm,
  columns=fullflexible,
  tabsize=2,
  showstringspaces=false,
  keepspaces=true
}

\AtBeginDocument{%
  }

\setcopyright{acmlicensed}
\copyrightyear{2026}
\acmYear{2026}
\acmJournal{TOSEM}
\acmVolume{1}
\acmNumber{1}
\acmArticle{}
\acmMonth{6}
\acmDOI{}

\makeatletter
\renewcommand{\@formatdoi}[1]{\ifx\@acmDOI\@empty\else\url{https://doi.org/#1}\fi}
\long\def\@makefntext#1{\noindent\hb@xt@1.8em{\hss\@thefnmark.\,}#1}
\makeatother

\begin{document}

\title[EnvPilot: Systematic Design and Evaluation of an Experience-Augmented Agent for Software Environment Setup]{EnvPilot: Systematic Design and Evaluation of an Experience-Augmented Agent for Software Environment Setup}


\author{Hanwu Chen}
\authornote{These authors contributed equally to this work.\quad
\textsuperscript{\dag}. For correspondence, please contact Yin Chen and Daoguang Zan.}
\affiliation{%
  \institution{Shenzhen Technology University}
  \city{Shenzhen}
  \country{China}
}
\email{2410263024@mails.szu.edu.cn}

\author{Hanyu Lin}
\authornotemark[1] 
\affiliation{%
  \institution{Shenzhen University}
  \city{Shenzhen}
  \country{China}
}
\email{2410104033@mails.szu.edu.cn}

\author{Zhanjiang Yang}
\affiliation{%
  \institution{Shenzhen Technology University}
  \city{Shenzhen}
  \country{China}
}
\email{2410263030@mails.szu.edu.cn}

\author{Linhao Zhang}
\affiliation{%
  \institution{Shandong University}
  \city{Shandong}
  \country{China}
}
\email{linhaoz2022@163.com}

\author{Aoyan Li}
\affiliation{%
  \institution{Huazhong University of Science and Technology}
  \city{Wuhan}
  \country{China}
}
\email{247736621@qq.com}

\author{Jinxi Li}
\affiliation{%
  \institution{Shenzhen Technology University}
  \city{Shenzhen}
  \country{China}
}
\email{ljjinxi@sztu.edu.cn}

\author{Meng Li}
\affiliation{%
  \institution{Shenzhen Technology University}
  \city{Shenzhen}
  \country{China}
}
\email{limeng2@sztu.edu.cn}

\author{Yin Chen\textsuperscript{\dag}}
\affiliation{%
  \institution{Shenzhen Technology University}
  \city{Shenzhen}
  \country{China}
}
\email{chenyin@sztu.edu.cn}

\author{Daoguang Zan\textsuperscript{\dag}}
\affiliation{%
  \institution{Institute of Software, Chinese Academy of Sciences}
  \state{Beijing Shi}
  \country{China}
}
\email{daoguang@iscas.ac.cn}

\renewcommand{\shortauthors}{Chen et al.}


\authorsaddresses{%
Hanwu Chen, Shenzhen Technology University, China;
Hanyu Lin, Shenzhen University, China;
Zhanjiang Yang, Shenzhen Technology University, China;
Linhao Zhang, Shandong University, China;
Aoyan Li, Huazhong University of Science and Technology, China;
Jinxi Li, Shenzhen Technology University, China;
Meng Li, Shenzhen Technology University, China;
Yin Chen, Shenzhen Technology University, China;
Daoguang Zan, Institute of Software, Chinese Academy of Sciences, China.\\
*For correspondence, please contact Yin Chen
(\href{mailto:chenyin@sztu.edu.cn}{chenyin@sztu.edu.cn})
and Daoguang Zan
(\href{mailto:daoguang@iscas.ac.cn}{daoguang@iscas.ac.cn}).%
}

\begin{abstract}
Environment Setup is a critical yet complex task in software engineering that relies heavily on expert knowledge. Existing automated environment setup methods lack the ability to accumulate experience from past execution trajectories and to evolve over time. As a result, their performance is limited because they often perform redundant exploration, ignore useful past solutions, and fail to generalize across diverse software ecosystems. \addedchw{We present the systematic design and empirical validation of EnvPilot, an experience-augmented agent that operationalizes trajectory-derived experience reuse for software environment setup.} EnvPilot maintains an expandable Trajectory-Derived Memory (TDM), initialized with 667 high-quality experiences. It systematically transforms implicit knowledge from historical execution trajectories into structured experience and retrieves the most relevant guidance during task execution through the Context-aware Retrieval mechanism. This enables EnvPilot to combine multiple validated setup strategies, providing more precise and detailed guidance than methods that rely solely on static project files or web retrieval. To evaluate EnvPilot, we construct AES-Bench, a multilingual benchmark of 112 real-world GitHub instances across 9 programming languages. Experiments show that EnvPilot achieves a new state-of-the-art (SOTA) with a 75.00\% Pass@1 success rate while reducing reasoning costs. Our empirical study shows that both the structured experience representation and the Context-aware Retrieval mechanism are essential. 
\end{abstract}



\received{30 November 2025}
\received[revised]{5 June 2026}
\received[accepted]{7 June 2026}

\maketitle

\section{Introduction}
Automated environment setup is an essential component of software development~\cite{envbench, executionagent, autoinstallinpy, hu2025repo2runautomatedbuildingexecutable}.
Tasks such as code execution, testing, and deployment all depend on a correct, stable, and reproducible runtime environment~\cite{yang2024acecode}. 
Likewise, with the growing use of Large Language Models (LLMs) in software engineering, the demand for reliable automated setup has increased~\cite{kovrigin2025piper}. 
\addedfinal{Automated environment setup is essential for AI-driven tasks. Specifically, high-quality datasets and Reinforcement Learning (RL) agents rely on code execution. For example, when we build training datasets~\cite{swebench,mswebench,swerebench,swefactory}, we must set up the environment to verify the code. If the environment fails, we cannot check if the code is correct. Similarly, in RL tasks, the agent needs to run code to get reward~\cite{wei2025swerl,rlcoder,zeng2025skywork}. A stable environment ensures that the agent learns from the code rather than from setup errors. Therefore, a reliable setup tool is essential infrastructure for these fields, ultimately advancing next-generation coding models}. 

Existing methods mainly rely on static information, such as project files and documentation, or use web retrieval. They also treat each setup task in isolation~\cite{autoinstallinpy, swefactory, executionagent, swebenchgoeslive, hu2025repo2runautomatedbuildingexecutable}. 
As a result, they often repeat the same mistakes, leading to wasted computational resources and limited learning across different setup tasks. 

Experience reuse offers a promising way to address these problems. It means learning from past attempts and applying that knowledge to new tasks. When knowledge from real usage is collected and applied, later work becomes more effective. Systems that learn from past fixes can suggest next steps beyond a developer's first query~\cite{kolodner1992introofcbr, zhang2023ecoassistant, tang2025agentkb, chen2025swe-exp}. 
At the macro level of the organization, experience becomes knowledge, which is then applied through tools and routines, ultimately lifting the next round of work. Context shapes what is learned, and stored knowledge reshapes later work~\cite{argote2011organizational}. 
Reusable and structured experience is therefore a strong lever for better performance. For environment setup tasks specifically, experience reuse can also improve outcomes significantly. If prior attempts are turned into simple cases and retrieved by context, then we avoid issues such as repeating the same errors, missing the chance to use validated solutions, and burning unnecessary compute.

However, existing automated environment setup methods fall short in leveraging such experience effectively~\cite{autoinstallinpy,envbench, executionagent, hu2025repo2runautomatedbuildingexecutable}. They operate in isolation, repeat errors, and fail to generalize across languages and projects. Memory-augmented systems~\cite{wang2024memoryllm, packer2023memgpt}, retrieval-augmented generation~\cite{lewis2020retrieval}, and experience replay~\cite{feng2025get, liu2025contextual} are already active research areas. Many techniques are available, such as graph-based memory~\cite{anokhin2024arigraph}, hierarchical storage~\cite{packer2023memgpt} and multimodal structured memory~\cite{li2024optimus}. Existing software engineering research also explores experience memory for code agents~\cite{chen2025swe-exp,tang2025agent, guo2026evoconfig}. However, directly applying these
off-the-shelf memory systems to automated environment setup has three problems. Firstly, raw setup trajectories are lengthy and noisy, so storing or replaying them as they are uses up a lot of tokens and makes it difficult to map error messages to specific solutions. Secondly, fix steps must be combined across different toolchains, such as programming languages, package managers and build systems. Replaying a full trajectory is not possible in these cases. Thirdly, many errors can be anticipated in repository files, such as README, CI scripts and build configurations, before execution. Therefore, retrieval that only runs after a failure wastes attempts.

To address these challenges, \addedchw{we present the systematic design and empirical validation of EnvPilot, an experience-augmented agent tailored to software environment setup.}
EnvPilot systematically transforms raw execution trajectories into structured experiential knowledge. Specifically, it initializes with 667 high-quality experiences distilled from past setup trajectories, each represented in a reusable \textit{problem–solution–action} tuple. This design allows the agent to record not only the errors that occurred, but also the effective corrective strategies and executable commands. The resulting experience memory is inherently extensible; each new task that is executed adds to the validated knowledge in the repository, which reflects an increasingly diverse range of environments and obstacles over time.

During the environment setup process, EnvPilot employs the Context-aware Retrieval mechanism that proactively anticipates setup challenges. By analyzing the codebase in detail, it constructs a clear and structured project profile (e.g., language versions, dependency managers, and build commands) and generates multiple hypothetical queries that model potential failure points. These queries probe the Trajectory-Derived Memory (TDM) from different perspectives, including conflicts, toolchain incompatibilities, or system-level mismatches. A multi-query retrieval strategy with Reciprocal Rank Fusion~\cite{cormack2009rrf} then consolidates the most relevant prior cases, ensuring that retrieved knowledge is not only semantically aligned with the current repository but also pragmatically useful for the environment setup process. The selected experiences are seamlessly injected into the agent's reasoning loop, guiding it to effectively leverage proven solutions.
In this way, EnvPilot transforms the automated environment setup process into a ``Learn–Retrieve–Act–Evaluate'' (LRAE) cycle: it learns from trajectories, retrieves relevant experiences, acts through containerized execution, and evaluates. \addedfinal{To ensure rigorous and scalable assessment, EnvPilot adopts an automated LLM-as-a-Judge mechanism aligned with standardized metrics~\cite{hu2025repo2runautomatedbuildingexecutable}, verifying setup success based on test lifecycle completeness rather than simple exit codes.} Each cycle enriches the TDM with new structured experiences. The agent evolves through continuous accumulation and refinement of experiences. This process shifts the environment setup paradigm from static and isolated reasoning to dynamic and trajectory-derived learning. EnvPilot can therefore generalize across diverse projects and languages. At the same time, it improves efficiency and robustness in a steady and measurable way. 

To systematically evaluate EnvPilot, we construct AES-Bench, a multilingual benchmark of 112 real-world GitHub instances across 9 programming languages. We also conduct extensive experiments using open-source LLMs. Results show that EnvPilot achieves an outstanding success rate of 75.00\% on AES-Bench, outperforming the existing method. Ablation studies show that both the TDM and the Context-aware Retrieval mechanism contribute to performance gains, and their combination yields the most significant improvement.
In summary, our main contributions are as follows.
\begin{enumerate}
    \item \textbf{EnvPilot.} \addedchw{We present the systematic design and empirical validation of EnvPilot, an experience-augmented agent tailored to automated software environment setup. EnvPilot operationalizes trajectory-derived experience reuse through two task-specific components: the TDM, which converts raw execution trajectories into reusable knowledge,} and the Context-aware Retrieval mechanism that analyzes project context and selects relevant prior experiences. \addedchw{Together, these designs enable EnvPilot to reuse validated setup knowledge across projects, reduce redundant exploration, and improve environment setup reliability.}
    \item \textbf{AES-Bench.} We construct the AES-Bench, which covers 112 real-world GitHub instances across 9 programming languages. Each instance is version-locked, contains executable tests, and reflects authentic build and dependency practices. This benchmark offers a rigorous way to evaluate automated environment setup methods. 
    \item \textbf{Extensive empirical study.} We evaluate EnvPilot and representative baselines on 
    AES-Bench. EnvPilot achieves the success rate of 75.00\%, the highest among all compared        
    methods. We also report ablation studies, an analysis of retrieval size, a failure taxonomy, and a 
    stability analysis.
\end{enumerate}
\section{Problem Statement}

\addedfinal{The goal of this work is to tackle the challenge of automated environment setup in modern software engineering. Such tasks require preparing a fully reproducible runtime environment so that a given project can be built smoothly and its tests reliably executed as expected.}
Here, we define an agent as an automated environment setup system equipped with the TDM. The agent observes a software project, retrieves relevant experiences from TDM, executes environment setup steps, and evaluates results through a closed-loop process.

\textbf{Problem definition.} The task is to set up the correct runtime environment for a given software project, specified by its repository and version. In this paper, the agent addresses this problem with the aid of the TDM, which provides structured experiential knowledge from past attempts and facilitates the reuse of validated solutions.

\textbf{Expected output.} \addedfinal{The agent configures the runtime environment inside a provided container and executes the project's test suite. The output includes (i) the executable commands sequence, including both successful and failed commands, (ii) the execution trajectory, and (iii) the test logs.}

This problem involves three key challenges:
\begin{enumerate}
    \item \textbf{Experience representation:} transforming raw and unstructured execution trajectories into clear and reusable structured memory entries  that retain critical environment setup details.
    \item \textbf{Precise retrieval:} identifying the most relevant experiences from a large memory to guide new projects.
    \item \textbf{Automated evaluation:} verifying environment setup outcomes reliably and without manual intervention, while ensuring evaluation precision meets preset requirements.
\end{enumerate}

\section{Method}
\begin{figure*}[ht]
  \centering
  \includegraphics[width=\textwidth]{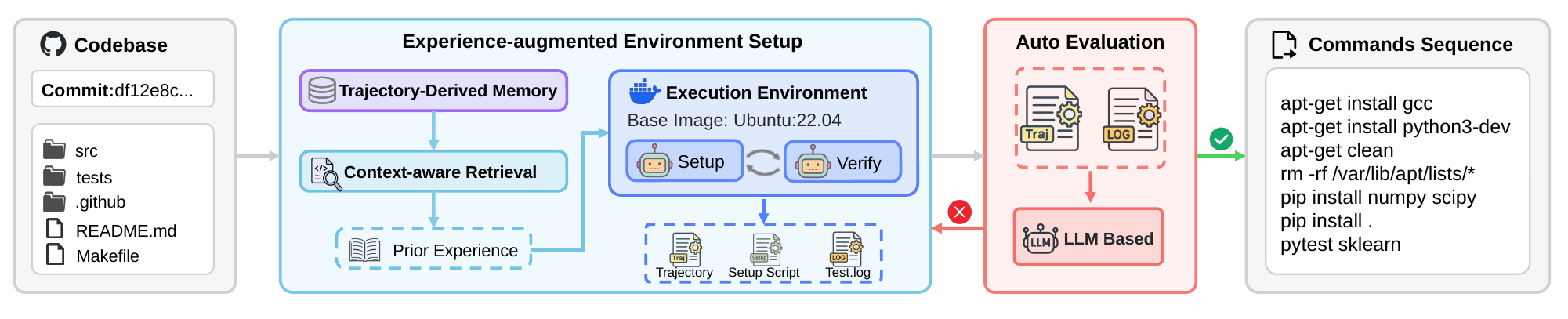}
  \caption{Overview of EnvPilot, an automated environment setup agent with experience memory and the Context-aware Retrieval mechanism. It follows a ReAct-style paradigm and executes tasks through an LRAE cycle.}
    \vspace{-1.5em}
  \label{fig:framework}
\end{figure*}

\subsection{Overview}

As shown in Fig.~\ref{fig:framework}, EnvPilot utilizes the TDM to store experiential knowledge from past execution trajectories, enabling the agent to perform environment setup tasks more effectively. The process by which the agent improves its environment setup skills over time follows a ``Learn–Retrieve–Act–Evaluate'' cycle:  

\begin{enumerate}
    \item \textbf{Learn}: Initially, the agent uses historical execution trajectories to build a clear and structured knowledge base in the TDM. These trajectories contain practical insights into common environment setup issues and solutions, forming the solid foundation for future tasks.
    \item \textbf{Retrieve}: When a new environment setup task begins, the agent first analyzes the current project's context. It then retrieves the most relevant experiences using a context-aware mechanism. This ensures that the agent learns from previous similar tasks and adapts past solutions to the new automated environment setup task.
    \item \textbf{Act}: The agent performs the environment setup inside an isolated container, using the guidance retrieved from the TDM. This ensures a clean environment for the environment setup process, minimizing external disruptions that could cause setup failures or inconsistencies.
    \item \textbf{Evaluate}: After completing the task, \addedfinal{the agent evaluates the outcome using an automated LLM-based evaluation mechanism. } This step verifies if the environment setup is successful or if adjustments are needed. New insights or corrections will be added to the TDM to improve future environment setup.
\end{enumerate}
\addedfinal{This cycle allows EnvPilot to continuously update the TDM. It learns from every new task and refines its behavior for future executions.}

\subsection{Trajectory-Derived Memory}
\label{sec:tdm}
\subsubsection{Experience Representation}
\label{sec:exp_rep}
\begin{figure}[!htbp]
  \centering
  \includegraphics[width=0.75\textwidth]{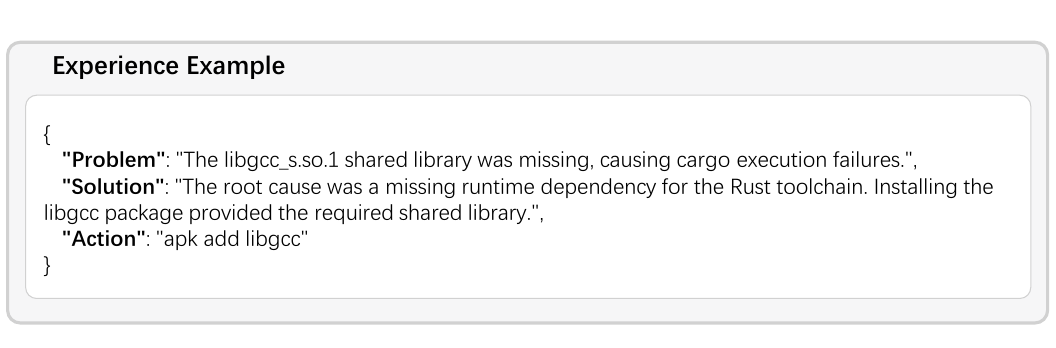}
  \caption{An example in our Trajectory-Derived Memory.}
  \label{fig:memory_example}
\end{figure}

Although a complete execution trajectory of an automated environment setup task contains extensive details, its length and lack of focus make it difficult to reuse directly. To address this, we design a structured representation that captures essential abstract patterns in a concise and reusable form.
Each experience entry is organized as a tuple with three components: the problem, which provides a natural language description of the encountered error; the solution, which summarizes the corrective strategy at a conceptual level; and the action, which specifies executable commands or script fragments that can be applied directly by the agent.  
An example of this representation is shown in Fig.~\ref{fig:memory_example}. In this case, the problem is described as the absence of the \texttt{libgcc\_s.so.1} shared library, which caused execution failures during Rust's \texttt{cargo} build process. The corresponding solution identifies the underlying cause as a missing runtime dependency in the Rust toolchain and indicates that installing the \texttt{libgcc} package resolves the issue. Finally, the action provides a concrete command, \texttt{apk add libgcc}, which can be executed to install the required library.  
This representation brings several advantages. It facilitates efficient matching between new problems and historical cases in a consistent manner, ensures that the retrieved knowledge is immediately executable for practical use, and simplifies the overall organization and maintenance of TDM to support long-term scalability.

\subsubsection{Memory Construction}
\label{sec:memory_construction}
The raw data for TDM originates from the execution trajectories of agents performing automated environment setup tasks in Multi-SWE-Bench~\cite{mswebench}, which provides realistic and challenging software engineering scenarios spanning multiple programming languages and dependency managers. \addedfinal{In total, we processed 781 raw execution trajectories, resulting in 667 distilled experience entries (an 85.4\% retention rate). Such diversity ensures that the extracted knowledge is both representative and generalizable. By processing these trajectories, we construct the initial knowledge base of TDM, designed to support the transfer and reuse of experiential knowledge across projects. The construction of TDM is highly automated, requiring only a one-time initialization of few-shot templates. This process unfolds in three stages: }

First, we conduct data collection and aggregation, gathering large-scale execution trajectories from historical tasks. At this stage, we retain both successful and failed cases without filtering. Successful trajectories reveal effective pathways under certain conditions, while failures expose common pitfalls and the causality between actions and negative outcomes.
\addedfinal{Secondly, we define high-quality entries based on three core criteria: (i) \textbf{Generality}: problem descriptions are refined to be repository-agnostic to ensure broad transferability; (ii) \textbf{Completeness}: each entry captures the full context of a bottleneck; and (iii) \textbf{Accuracy}: a strict factuality policy is enforced, recording only empirically proven solutions. To achieve these criteria, we implement a Chain-of-Thought (CoT) analysis pipeline that systematically identifies environment-related failures, performs deep root-cause analysis, and extracts validated interventions. We curate a small set of expert-annotated ``seed'' examples (typically $k=3 \sim 5$) reflecting this multi-step reasoning to serve as in-context learning templates. Each example defines a mapping from a raw trajectory to a structured \texttt{<problem, solution, action>} triplet. Within this triplet, the \texttt{problem} is a concise, agnostic description; the \texttt{solution} provides a detailed explanation of the root cause; and the \texttt{action} records the specific executable command (e.g., \texttt{apt-get install}). This manual curation represents a fixed, one-time cost; once the prompt is established, the system leverages an LLM to autonomously abstract implicit insights from verbose trajectories, scaling to thousands of cases without further human intervention.}
\addedfinal{Finally, we use DeepSeek-R1 as a reasoning engine to perform reasoning-based distillation. After the predefined analysis steps of problem identification, root cause analysis and solution extraction, the engine filters out noisy environmental signals and converts implicit insights into standardized JSON objects. To ensure methodological soundness, we conducted a manual audit on a random 5\% sample (approx. 34 entries), confirming that the entries accurately reflected original environment challenges and provided technically sound interventions.}
Formally, a given trajectory is defined as:
\begin{equation}
    \text{Traj} = \{\, o_1,\, t_1,\, a_1,\, \dots,\, o_{t},\, t_{t},\, a_{t} \,\}
\end{equation}
where $o$ denotes an \emph{observation}, $t$ a \emph{thought}, and $a$ an \emph{action}. Then an LLM, prompted with a context $C$ (containing guiding principles and few-shot in-context learning examples),  is employed to generate a structured experience $E$:
\begin{equation}
    E = \{\text{problem}, \text{solution}, \text{action}\} \leftarrow LLM(\text{Traj}, C).
\end{equation}
\addedfinal{In this formulation, the \texttt{problem} dimension encapsulates the symptom and root cause, \texttt{solution} provides the abstract corrective strategy, and \texttt{action} records the validated command sequence derived from the trajectory.}

\addedfinal{Instead of raw trajectories, TDM stores abstract patterns. This structure is concise and efficient. It allows the agent to match and reuse past solutions easily.}
To illustrate this process, consider a trajectory where the agent tries to set up a Python project and encounters a \texttt{ModuleNotFoundError} for numpy. The raw trajectory may contain many tool calls, partial logs, and repeated attempts. We do not store this raw trace in the memory. Instead, the LLM reads the entire sequence and answers three simple questions: what is the main symptom, what is the most likely root cause, and what strategy fixed the problem. In this example, the symptom is the \texttt{ModuleNotFoundError}, the root cause is a missing dependency in a clean environment, and the strategy is to install a compatible version of numpy with the package manager. The final experience entry is short, but it keeps the key information that later tasks can reuse.
This abstraction has two important effects. First, it normalizes many surface forms of similar problems into a small set of patterns. Different trajectories may show slightly different log messages for import errors, link errors, or version conflicts, but they often share the same root cause and strategy. The memory representation makes these common patterns explicit. Second, it reduces noise for the agent. During environment setup, the agent no longer needs to scan long logs from past runs. It can focus on a few short entries that clearly state what went wrong, why it happened, and what action worked in practice.

\begin{figure*}[ht]
    \centering
    \includegraphics[width=\textwidth]{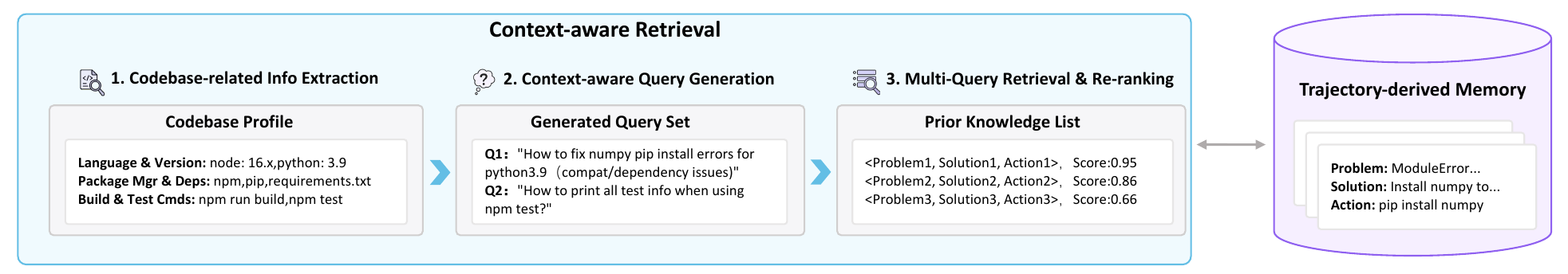}
    \caption{Overview of the Context-aware Retrieval mechanism.}
    \label{fig:context-aware_retrieval}
    \vspace{-1em}
\end{figure*}

\subsection{Experience-augmented Automated Environment Setup}
\subsubsection{Context-aware Retrieval}
We design the Context-aware Retrieval mechanism to find the most relevant content from large-scale historical experience. This module enables more targeted, context-adaptive experience reuse. As shown in Fig.~\ref{fig:context-aware_retrieval}, the Context-aware Retrieval mechanism consists of three steps.

\textit{Step1: Codebase-related Info Extraction.} 
The initial step involves a deep analysis of the target codebase to construct a structured ``codebase profile''. \addedfinal{The objective is to identify the precise technical components and commands required to configure the environment directly within a fresh Linux container. LLMs are used to parse project documentation (e.g., README.md), configuration files (e.g., pyproject.toml), and CI/CD files (e.g., ci.yaml) to extract essential technical details, such as programming language version, dependency management tools, and core build and test commands. Crucially, the extraction process prioritizes identifying system-level dependencies, such as those found in CI workflow steps or embedded within Dockerfile RUN commands, while ignoring infrastructure-level commands, such as Docker build or run. This information is then consolidated into a unified, structured representation that follows a specific schema. This schema serves as the foundational context for the subsequent step.}

\textit{Step2: Context-aware Query Generation.}
After generating the codebase profile, the system constructs a set of queries. Instead of \addedfinal{simple keywords, these queries are formulated as intent-aligned, hypothetical problem statements that anticipate potential obstacles within a fresh Linux container environment. This strategy leverages the pre-trained knowledge of modern LLMs regarding common repository configurations. These models are pre-trained on large datasets, including GitHub repositories and setup tasks, enabling them to identify latent setup patterns and version-specific pitfalls. We therefore hypothesize that such models can effectively generate reliable problem-oriented queries based on the codebase profile. By incorporating context-rich details such as OS versions, library constraints, and system-level build tools (e.g., apt-get dependencies, gcc, make), the generated queries focus on the internal configuration process while strictly excluding external containerization orchestration. For example, if the profile specifies Python 3.9 and PyTorch 1.8, the LLM generates descriptive natural language queries such as: ``How to resolve compatibility issues between PyTorch 1.8 and Python 3.9 in a Linux environment?'' Designed to be context-sensitive, these queries serve as semantic probes to identify environment setup obstacles within the target codebase proactively. The resulting set is then used to retrieve relevant memory fragments from TDM to guide later reasoning.}

\textit{Step3: Multi-Query Retrieval and Re-ranking.}  
To this end, we employ a multi-query retrieval mechanism. Specifically, given a set of generated queries $Q = \{q^1, q^2, \dots, q^n\}$ from the previous step and our proposed TDM $M$ consisting of experience entries $e_k = \{\text{problem}, \text{solution}, \text{action}\}$, we proceed as follows.
For each query $q^i \in Q$, we first embed it into a vector representation using an embedding model. We then compute the cosine similarity between the vector of $q_i$ and the \textit{problem} vector of every experience entry $e_k \in M$, resulting in a candidate set $C_i$. Each $C_i$ constitutes a ranked list in descending order of similarity scores: 
\begin{equation}
    C_i = \{ \langle e_i^1, s_i^1 \rangle, \langle e_i^2, s_i^2 \rangle, \dots, \langle e_i^N, s_i^N \rangle \}
\end{equation}
where $e_i^j$ denotes the experience with the $j$-th highest similarity score for query $q_i$, $s_i^j$ represents the corresponding cosine similarity score, and $N$ represents the total number of retrieved candidates per query. The candidate sets from all queries are then aggregated and re-ranked using Reciprocal Rank Fusion~\cite{cormack2009rrf}. The score for an experience entry $e_k$ is defined as equation \ref{eq:rrf} for clarity:
\begin{equation}
\label{eq:rrf}
\text{Score}(e_k) = \sum_{i=1}^{n} \frac{1}{\text{rank}_i(e_k) + p}
\end{equation}
where $\text{rank}_i(e_k)$ denotes the rank of $e_k$ in candidate set $C_i$, $n$ is the total number of queries, and $p$ is a smoothing constant (typically set to 60) to mitigate the impact of high ranks. The final retrieved memory set is obtained by sorting all candidates in descending order of their scores. The top-ranked memory, along with the codebase profile, is injected into the agent's context to provide executable and verifiable guidance for automated environment setup tasks.

In practice, the three steps in our retrieval pipeline work together in a natural way. Suppose the target repository is a JavaScript project that contains both \texttt{npm} and \texttt{pnpm} lock files, and the tests are triggered by \texttt{npm} test. The codebase profile records this mixed toolchain and the main test command. Based on this profile, the LLM generates several queries. Some queries ask how to handle lock file conflicts between \texttt{npm} and \texttt{pnpm}. Other queries focus on common failure patterns when \texttt{npm} test stops early or hides test output. Each query then retrieves a slightly different subset of experiences, for example past cases where node versions or package managers were misaligned. After Reciprocal Rank Fusion, we obtain a final ranked list that combines all these signals. The top entries tend to be robust solutions that have helped many similar trajectories.
We also observe a trade-off between the number of retrieved items and the stability of the agent. A larger retrieved set can provide more guidance, but it also increases context length and may introduce irrelevant commands that distract the agent. In EnvPilot, we therefore keep the number of retrieved items moderate and rely on the ranking step to surface the most helpful experiences first. This design choice balances guidance and stability, which helps the agent act in a steady and dependable way.

\subsubsection{Containerized Execution}

After retrieving the most relevant experience entries, EnvPilot proceeds to the environment setup phase. To guarantee reproducibility and security, each task is executed within an isolated containerized environment. Containers provide a controlled runtime that eliminates interference from residual states of prior tasks, ensuring that every run begins from a clean state. This design also strengthens security by confining potential failures or side effects to the container itself, thereby protecting the host system.
During the process, all interactions between the agent and the environment are comprehensively recorded, including observations, reasoning steps, and actions. These records constitute a complete execution trajectory that serves two purposes. First, they offer clear and verifiable evidence of how retrieved experiences are applied in practice, enabling both automated evaluation and human inspection. Second, the recorded trajectory supports the continuous and incremental learning of the TDM. In addition, the system also generates a separate test log that captures the outputs produced during test execution. This log file further supports reliable evaluation and later analysis.
Once a task is completed, EnvPilot analyzes the trajectory and updates the TDM accordingly. Successful cases enrich the TDM with validated strategies and executable actions, while failures contribute valuable negative experiences that reveal causal links between actions and negative results. Through this iterative feedback loop, EnvPilot progressively refines its knowledge base and adapts to the diversity of real-world projects.

\subsection{\addedfinal{Automated Evaluation Framework}}
\label{sec:hybrid_eval}
\addedfinal{To ensure rigorous and scalable feedback, we employ a pure LLM-as-a-Judge mechanism. Following the standard metric in Repo2Run~\cite{hu2025repo2runautomatedbuildingexecutable}, we adopt the Environment Configuration Success Rate (ECSR). A critical challenge in this domain is distinguishing between environment failures and inherent code bugs. Therefore, our evaluation core principle is to judge the \textit{integrity of the environment}, not the correctness of the repository's code logic.}

\addedfinal{\textit{LLM-as-a-Judge Implementation.}
We employ DeepSeek-V3 as the judge, driven by a specialized prompt that enforces strict differentiation between failure types. The model analyzes the full execution trajectory, specifically focusing on the output of the final executed command to determine the ultimate environment state. The judgment logic is governed by three key criteria:}
\begin{enumerate}
    \item \addedfinal{\textbf{Distinction of Failure Types:} The judge distinguishes between \textit{Assertion Failures} (e.g., \texttt{assert 1==2}) and \textit{Environment Errors} (e.g., \texttt{ImportError}, \texttt{Segmentation Fault}). Assertion failures indicate that the environment is successfully configured and capable of running tests, thus counting as \texttt{PASSED}. Conversely, runtime crashes, missing dependencies, or compilation errors are marked as \texttt{FAILED}.}
    
    \item \addedfinal{\textbf{Test Lifecycle Completeness:} A configuration is deemed successful if and only if the test process completes a full lifecycle. This requires the test runner to start, execute, and gracefully exit, producing a valid final report or summary (e.g., ``5 passed, 2 failed''). Processes that abort unexpectedly or produce incomplete logs are rejected.}
    
    \item \addedfinal{\textbf{Compilation and Execution:} For compiled languages, the source code must compile successfully. For all languages, the agent must explicitly trigger the test command. Cases where dependencies are installed but no tests are executed are classified as failures.}
\end{enumerate}

\addedfinal{This prompt design ensures that the agent is rewarded for establishing a runnable environment, even if the underlying repository contains failing tests, thereby aligning the metric with the practical goal of environment setup.}

\section{Benchmark}
\label{sec:benchmark}
To rigorously evaluate the agent's ability in repository-level environment setup and test execution, we introduce AES-Bench. Existing benchmarks show clear limitations for such tasks. One type, represented by EnvBench~\cite{envbench}, determines success through zero exit codes and static analysis tools. This approach is restricted to specific languages (only Python and Java) and ignores the critical step of test execution. Without test execution, the evaluation cannot verify whether the configured environment actually supports building and running the project. Another type, represented by ExecutionAgent~\cite{executionagent}, focuses on execution but lacks reproducibility, since it is not tied to specific code commit versions. As a result, evaluation results drift over time, which makes it difficult to obtain stable measurements and prevents fair comparison across different methods.
To fill this gap, AES-Bench follows three core principles: version locking, full test maintenance, and multilingual coverage. First, to ensure perfect reproducibility, each instance is bound to a specific commit SHA. Second, to reflect real workflows, the agent must not only configure and build the project but also execute at least one native test suite as a success signal. Finally, to support broad applicability across diverse toolchains, the benchmark covers 112 task instances spanning 9 programming languages.

\subsection{Dataset Construction}

We construct AES-Bench with clear and strict conditions to ensure diversity and reproducibility. The selection of repositories is based on the following three conditions:

\begin{enumerate}
\item \textbf{Repository popularity and stability.}
Repositories must have at least 1000 stars on GitHub. This guarantees that the projects are widely used and well-maintained.

\item \textbf{Sustained development activity.}  
Each repository must show continuous commits and active issue resolution for at least six months before collection. This ensures that the projects are alive and supported by communities.  

\item \textbf{Testing infrastructure.}  
Every candidate repository must provide executable tests, such as CI/CD pipelines, unit tests, or integration tests. This is necessary for the automated evaluation of the environment setup.  

\end{enumerate}

\begin{figure}[!htbp]
  \centering
  \includegraphics[width=0.75\textwidth]{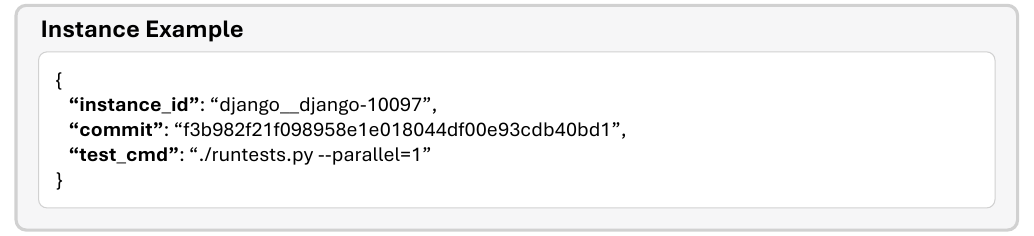}
  \caption{An instance example in AES-Bench.}
  \label{fig:instance_example}
\end{figure}

After applying the three conditions, we obtain 99 qualified repositories. From these repositories, \addedfinal{we systematically select candidates following the criteria to ensure diversity across languages, dependency management tools, and build systems.} It is important to note that one repository can have multiple versions. One key factor is that environment differences between a repository’s major versions can be significant.  For this reason, different commits from the same repository may be included as separate instances in our dataset. Following this process, we finally construct 112 instances covering 9 mainstream programming languages. \addedfinal{To ensure reproducibility, the complete list of all instances with their commit information is provided in Appendix~\ref{appendix:dataset_instances}. Each instance is represented by three fields: an identifier (\texttt{instance\_id}), a commit hash (\texttt{commit}) and a test command (\texttt{test\_cmd}). The identifier is composed of the organization name, the repository name, and a pull request number, which naturally links to a specific commit. An example is shown as Fig.~\ref{fig:instance_example}.} This structure offers a compact and consistent way to reference each repository. Every instance is tied to a fixed commit SHA, which ensures reproducibility and stable evaluation results. The dataset therefore reflects diversity in language and scale, captures authentic practices of dependency management and build systems, and provides a realistic basis for evaluating automated environment setup methods.

\addedfinal{When constructing AES-Bench, we strictly enforced data separation. The repositories in AES-Bench are completely disjoint from the repositories used for TDM. We filtered the data by repository name. If a repository appeared in the memory construction set, we removed it from the benchmark. Therefore, the evaluation data consists of entirely unseen projects. Please refer to Tab.\ref{tab:deduplicated_github_repos} for memory sources and Appendix~\ref{appendix:dataset_instances} for the full benchmark list.} This separation prevents direct data leakage between memory construction and evaluation. As a result, AES-Bench tests whether abstracted experience can transfer to new projects with different dependency graphs and build scripts, rather than whether the agent can memorize complete solutions from past runs. At the same time, the selection rules ensure that each instance is realistic enough to reflect real engineering practice while still being reproducible in a clean Ubuntu 22.04 environment.

\subsection{Dataset Statistics}
AES-Bench covers 9 mainstream programming languages: C, C++, Go, Java, JavaScript, PHP, Python, Rust, and TypeScript. For each language, we select 12 or 13 representative instances to ensure both diversity and balance. As shown in Tab.~\ref{tab:repo-stats-compact}, projects in different languages vary greatly in popularity and scale. Go, JavaScript, and Rust projects have higher average star counts, showing more vigorous community activity. Java projects contain the largest average number of files, which reflects their complex engineering structure. C and C++ projects also show large scales, with nearly 6000 files on average, and often rely on traditional build tools such as \texttt{make} and \texttt{cmake}. All projects tend to use mature CI and broad test suites. This raises the bar for automated environment setup. The agent must respect pinned toolchain versions, run project-specific scripts, and install native system packages when needed. Ecosystem differences lead to distinct failure modes. These properties make AES-Bench a realistic stress test for dependency resolution, version management, and reproducibility.

\begin{table}[h]
\centering
\caption{Statistics for repositories by dominant language. Tool adoption shown as percentages (count/total).}
\label{tab:repo-stats-compact}
\setlength{\tabcolsep}{8pt}
\small
\begin{tabularx}{\linewidth}{lcccX}
\toprule
Language & \#Inst. & Avg.\ Stars & Avg.\ Files & \RaggedRight Tool Adoption \\
\midrule
C          & 12 & 24966 & 5994 & cmake 66.67\% (8/12), make 91.67\% (11/12), meson 33.33\% (4/12) \\
C++        & 13 & 26350 & 5668 & cmake 92.31\% (12/13), make 92.31\% (12/13), conan 7.69\% (1/13) \\
Go         & 13 & 57473 & 1316 & go~mod 100.00\% (13/13) \\
Java       & 13 & 11419 & 7255 & maven 92.31\% (12/13), gradle 30.77\% (4/13) \\
JavaScript & 12 & 46075 & 1013 & npm 66.67\% (8/12), yarn 16.67\% (2/12), node 25.00\% (3/12) \\
PHP        & 12 & 19760 & 1332 & composer 100.00\% (12/12) \\
Python     & 12 & 29506 & 1597 & pip 91.67\% (11/12), conda 8.33\% (1/12) \\
Rust       & 12 & 38178 & 1715 & cargo 100.00\% (12/12) \\
TypeScript & 13 & 35136 & 3189 & npm 30.77\% (4/13), pnpm 46.15\% (6/13), yarn 38.46\% (5/13) \\
\bottomrule
\end{tabularx}
\end{table}

We also analyze the use of build systems and dependency management tools. In C and C++ projects, \texttt{make} and \texttt{cmake} dominate, with more than 90\% adoption in C++. Go projects all use \texttt{go mod}, showing a unified ecosystem. Java projects mainly adopt \texttt{maven}, with some using \texttt{gradle}. JavaScript and TypeScript projects show diverse choices, including \texttt{npm}, \texttt{yarn}, and \texttt{pnpm}, which require the agent to resolve conflicts and priorities. Python projects rely primarily on \texttt{pip}, with a few using \texttt{conda}. Rust projects uniformly use \texttt{cargo} and PHP projects all use \texttt{composer}. These findings show that our dataset is not only representative in language and project scale but also reflects real practices of build and dependency tools in the open-source community. In summary, some ecosystems, such as Go, Rust, and PHP, are standardized and easier for agents to set up. Others, such as JavaScript and C/C++, are more diverse and complex, which poses greater challenges for automatic environment setup. This provides a realistic and challenging basis for evaluating agents in real-world software engineering tasks.



\section{Experimental Setup}
\label{sec:experiments}
We design our experiments to answer seven research questions. Each question focuses on a key aspect of EnvPilot, including effectiveness, module contribution, evaluation reliability, experience usage, and real-world interpretability.

\begin{itemize}[leftmargin=12pt]
    \item \textbf{RQ1: Does EnvPilot improve success rate and efficiency in automated environment setup compared to existing methods?}
    We compare EnvPilot with SOTA automated environment setup agents on AES-Bench. We evaluate success rate, token cost, and interaction steps. This experiment shows whether EnvPilot can achieve higher environment setup success.

    \item \textbf{RQ2: How do the TDM and the Context-aware Retrieval mechanism contribute to EnvPilot’s performance?}
    We conduct ablation studies by removing or replacing individual components. We measure performance drops to understand how each module supports EnvPilot. This experiment verifies whether the structured memory and retrieval strategy are necessary and whether they provide stable benefits across different scenarios.
    \addedfinal{\item \textbf{RQ3: Is the automated LLM-as-a-Judge evaluation framework reliable compared to human experts' assessment?}
    We validate the reliability of our automated evaluation strategy by comparing its judgments against a human-annotated ground truth dataset. This dataset consists of 460 execution trajectories generated by DeepSeek-V3-0324 on AES-Bench. This experiment measures the agreement (Accuracy, Precision, Recall, F1) between the LLM judge and human experts to ensure the automated metric serves as a valid proxy for manual verification.}
    
    \item \textbf{RQ4: How does the number of retrieved experiences affect the effectiveness of environment setup?}
    We vary the number of retrieved experience items during environment setup. This experiment examines whether more experience always leads to better results.

    \item \textbf{RQ5: Are the gains from TDM stable across repeated runs under LLM nondeterminism?} We repeat the core
    comparison between EnvPilot with TDM and EnvPilot without TDM three times under the same settings. We report
    the mean and standard deviation of success rate and cost to assess stability.
    
    \item \textbf{RQ6: \addedfinal{In a real-world repository case study}, how does experience augmentation change the agent’s behavior?}
    We perform a case study to compare the trajectory of EnvPilot with and without experience in a controlled setting. This experiment shows how experience changes the agent’s decision process, improves its reasoning steps, and helps explain measurable improvements in success rate.
    
    \item \textbf{RQ7: What types of failures still occur, and what do these failures reveal about the current limits of automated environment setup?}
    We explain why some instances cannot be configured successfully by EnvPilot. It also allows us to study how different tools, runtimes, and system constraints interact with the agent’s decisions. By examining each failed case, we can identify common patterns, locate the main sources of errors, and clarify which parts of the setup process remain difficult for current agents. Such analysis provides guidance for future improvements in automated environment setup strategies.

\end{itemize}

\subsection{Baselines}
\label{subsec:baselines}

We compare EnvPilot against the following baselines:
\begin{itemize}[leftmargin=12pt]
    \item \textbf{SWE-Agent (Adapted for Environment Setup).}
    This baseline~\cite{yang2024sweagent} was originally designed for repository-level issue resolve on SWE-bench~\cite{swebench}. It interacts efficiently with the operating system through its proposed Agent-Computer Interface. For our study, we adapted it from code repair to automated environment setup, focusing on dependency installation and test suite execution.

    \item \textbf{RepoLaunch.}
    This baseline is the automated environment setup module in SWE-bench-Live~\cite{swebenchgoeslive}. It selects a suitable base image by key environment configuration files and employs a ReAct-style~\cite{yao2023react} agent to install dependencies and validate tests. By reducing manual effort and enhancing scalability, we consider it a strong baseline for setup tasks.
    
    \item \textbf{ExecutionAgent.}
    This baseline~\cite{executionagent} leverages meta-prompting to autonomously set up environments through an execution-feedback loop. It has achieved a high success rate in multilingual GitHub projects, showcasing its potential as a general automated environment setup agent.
    
    \item \addedfinal{\textbf{Repo2Run.}This baseline~\cite{hu2025repo2runautomatedbuildingexecutable} is an LLM-based agent designed to automate the creation of executable environments by iteratively building Docker images and running tests. Notably, it utilizes a specialized toolset tailored for Python. We adapted Repo2Run to align with the AES-Bench evaluation criteria. The original implementation stops after collecting tests. We modified the \texttt{runtest} and \texttt{poetryruntest} tools to execute tests immediately after collection. The agent now outputs full execution logs. Note that we still assess Repo2Run only on the Python subset of AES-Bench due to its inherent language constraints.
    }
    
    \item \addedfinal{\textbf{Installamatic.}
    This baseline~\cite{envbench} follows a non-agentic configuration approach. It first retrieves relevant documentation and then attempts to output a complete environment setup script. Unlike ReAct-style agents, it lacks feedback from environment to self-correct, serving as a representative for retrieval-augmented script generation methods.
    }
    
    \item \addedfinal{\textbf{Bash-based Agent.}
    Introduced in EnvBench~\cite{envbench}, this agent uses a ReAct loop for planning and bash execution. The original implementation relies on separate agents coupled with Python and Java. We adapted this approach to be generalizable. We replaced the language-specific prompt with a unified, language-agnostic prompt. The core bash execution and ReAct mechanism remain unchanged. This modification allows the baseline to function across the diverse programming languages in AES-Bench.
    }
\end{itemize}

\subsection{Evaluation Metrics}
\label{subsec:metrics}

\addedfinal{We evaluate agent performance along three dimensions: effectiveness, cost, and efficiency. The main metric is the Environment Configuration Success Rate (ECSR), aligned with recent studies~\cite{hu2025repo2runautomatedbuildingexecutable}. 
A task is considered successful strictly if the agent can configure the environment such that the project's tests execute to completion (i.e., a full test lifecycle is observed), regardless of whether specific test cases pass or fail. This distinguishes environment integrity from inherent code defects. 
To reflect resource usage and interaction efficiency, we also report the average number of interactions per task (\texttt{API Calls}), the average number of tokens consumed per task (\texttt{Token}), and the average monetary cost based on token usage per task (\texttt{Dollar Cost}).}

\subsection{Implementation Details}
\label{subsec:implementation}
We first describe how we construct the TDM and then explain the key parameter choices used during training and evaluation, for clarity and reproducibility.

\begin{table}[!ht]
    \centering
    \caption{Selected Automated Environment Setup Task Instances for TDM's Initial Stage.}
    \label{tab:deduplicated_github_repos}
    \begin{adjustbox}{max width=\linewidth}
    \begin{tabular}{lll}
        \toprule
        alibaba/fastjson2 & darkreader/darkreader & googlecontainertools/jib \\
        anuraghazra/github-readme-stats & elastic/logstash & grpc/grpc-go \\
        apache/dubbo & expressjs/express & iamkun/dayjs \\
        axios/axios & facebook/zstd & jqlang/jq \\
        BurntSushi/ripgrep & facebookresearch/hydra & Kong/insomnia \\
        catchorg/Catch2 & fasterxml/jackson-core & mockito/mockito \\
        clap-rs/clap & fasterxml/jackson-databind & mui/material-ui \\
        cli/cli & fasterxml/jackson-dataformat-xml & nlohmann/json \\
        conan-io/conan & fmtlib/fmt & nushell/nushell \\
        ponylang/ponyc & python/mypy & rayon-rs/rayon \\
        serde-rs/serde & sharkdp/bat & sharkdp/fd \\
        simdjson/simdjson & sveltejs/svelte & tokio-rs/bytes \\
        tokio-rs/tokio & tokio-rs/tracing & vuejs/core \\
        yhirose/cpp-httplib & zeromicro/go-zero & / \\
        \bottomrule
    \end{tabular}
    \end{adjustbox}
\end{table}

\textbf{TDM Construction.} We bootstrap the TDM with execution trajectories generated by SWE-Agent~\cite{yang2024sweagent} on automated environment setup tasks across diverse GitHub repositories (Tab.\ref{tab:deduplicated_github_repos}), including projects representative of benchmarks like Multi-SWE-Bench\cite{mswebench}.
These trajectories, spanning programming languages such as Python, C, Java, Go, along with varied frameworks and dependency managers, provide a broad sample of real-world software engineering challenges.
The curated trajectories are then transformed into structured experience entries \texttt{<problem, solution, action>} via a large reasoning model. In our implementation, we employ DeepSeek-R1-0528~\cite{guo2025deepseekr1} with its official parameters, resulting in 667 knowledge items distinct from the AES-Bench dataset and covering common environment setup scenarios. This variety equips the memory with reusable plans and diagnostic insights for effective support in novel environments.

\begin{figure}[tbp]
    \centering
    \includegraphics[width=0.72\linewidth]{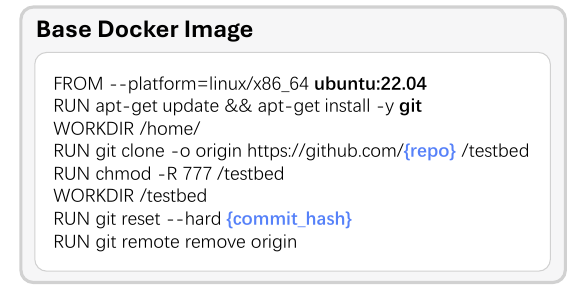} 
    \caption{Base image used for each automated environment setup task.}
    \label{fig:base_image}
\end{figure}

\textbf{Environment Setup.} EnvPilot integrates TDM into the SWE-Agent~\cite{yang2024sweagent} framework, which is already adapted for environment setup tasks. This enhances the capabilities of these tasks, providing a high-quality, expert-level experience.  For each task, we construct a base image (Fig.~\ref{fig:base_image}) and restrict the agent to a maximum of 50 interaction turns. 
The generative model is DeepSeek-V3-0324~\cite{liu2024deepseek}, with temperature set to 0.0 and top-p to 1.0 for deterministic outputs. Text embeddings are produced by Qwen3-Embedding-8B~\cite{zhang2025qwen3embedding}, which generates 4096-dimensional vectors. The retrieval module returns up to $k$ relevant knowledge items, with $k=10$ as the default. A summary of these configurations is provided in Tab.~\ref{tab:hyperparams}. \addedfinal{Note that while EnvPilot is designed with the capability for continuous learning, we froze the TDM during the evaluation. The memory state remained constant, containing only the initial 667 high-quality experience items. This ensures that the evaluation results are deterministic and independent of the execution order of the benchmark instances. To ensure the reliability of our results and mitigate the inherent randomness of LLMs, we adopted a rigorous evaluation strategy. For our proposed EnvPilot and the primary baseline adapted SWE-Agent, we conducted three independent runs for each instance. We report the median success rate (Pass@1) of these three runs in our main experimental results. This approach ensures that the performance gains are stable and not due to lucky generations. For other baselines, we report the results of a single run.}

\begin{table}[ht]
\centering
\small
\caption{Key hyperparameters and configuration for our experiments.}
\label{tab:hyperparams}
\begin{tabular}{llc}
\toprule
\textbf{Component} & \textbf{Parameter} & \textbf{Value} \\
\midrule
\multirow{4}{*}{Generative Model}
& Generative Model & DeepSeek-V3-0324~\cite{liu2024deepseek} \\
& Max Iterations & 50 \\
& Temperature & 0.0 \\
& Top-p & 1.0 \\
\midrule
\multirow{4}{*}{Memory Retriever}
& Vector Database & Milvus \\
& Embedding Model & Qwen3-Embedding-8B~\cite{zhang2025qwen3embedding} \\
& Embedding Dimension & 4096 \\
& Max Retrieved Items & 10 \\
\bottomrule
\end{tabular}
\vspace{-2em}
\end{table}

\section{Experiment Results}
\subsection{RQ1: Comparison with Baseline Methods for Environment Setup Performance}

\begin{table}[t]
  \centering
  \small
  \caption{Cross-benchmark performance comparison between AES-Bench and ExecutionAgent-Bench.}
  \label{tab:cross_benchmark_comparison}
  \resizebox{\textwidth}{!}{%
  \begin{tabular}{ll|lccc|lccc}
    \toprule
    \multirow{2}{*}{\textbf{LLM}} & 
    \multirow{2}{*}{\textbf{Method}} & 
    \multicolumn{4}{c|}{\textbf{AES-Bench (Ours)}} & 
    \multicolumn{4}{c}{\textbf{ExecutionAgent-Bench}} \\
    \cmidrule(lr){3-6}\cmidrule(lr){7-10}
    & & \textbf{ECSR (\%)} $\uparrow$ & \textbf{API} $\downarrow$ & \textbf{Tokens (K)} $\downarrow$ & \textbf{Cost (\$)} $\downarrow$
      & \textbf{ECSR (\%)} $\uparrow$ & \textbf{API} $\downarrow$ & \textbf{Tokens (K)} $\downarrow$ & \textbf{Cost (\$)} $\downarrow$ \\
    \midrule
    \multirow{7}{*}{GPT-4o-mini} 
        & ExecutionAgent & 23.21 \textcolor{gray}{(26/112)} & 44 & 643 & 0.105 & 18.00 \textcolor{gray}{(9/50)} & \textbf{31} & 480 & 0.079 \\
        & Repo2Run      & 25.00 \textcolor{gray}{(3/12)}   & 45 & 1273 & 0.193 & 27.27 \textcolor{gray}{(3/11)} & 43 & 1356 & 0.206 \\
        & Bash-Agent    & 20.54 \textcolor{gray}{(23/112)} & 33 & 380 & 0.059 
        & \underline{36.00} \textcolor{gray}{(18/50)} & 33 & 466 & 0.071 \\
        & RepoLaunch   & \underline{40.18} \textcolor{gray}{(45/112)} & 44 & 387 & 0.060 
        & {30.00} \textcolor{gray}{(15/50)} & {39} & {365} & {0.056} \\
        & SWE-Agent     & 24.11 \textcolor{gray}{(27/112)} & \underline{29} & \textbf{192} & \textbf{0.030}
        & 28.00 \textcolor{gray}{(14/50)} & 33 & \textbf{209} & \textbf{0.033} \\
        \rowcolor[gray]{0.95} &  \textbf{EnvPilot (Ours)} & \textbf{46.43} \textcolor{gray}{(52/112)} & \textbf{26} & \underline{214} & \underline{0.034}
        & \textbf{46.00} \textcolor{gray}{(23/50)} & \underline{32} & \underline{227} & \underline{0.036} \\
    \midrule
    \multirow{7}{*}{DeepSeek-V3} 
        & ExecutionAgent & 43.75 \textcolor{gray}{(49/112)} & \underline{28} & 420 & 0.130 & 36.00 \textcolor{gray}{(18/50)} & \underline{27} & 422 & 0.131 \\
        & Repo2Run  & 66.67 \textcolor{gray}{(8/12)}   & 20 & 1000 & 0.281 
        & 54.54 \textcolor{gray}{(6/11)} & 15 & 847 & 0.238 \\
        & Bash-Agent    & -- & -- & -- & -- & -- & -- & -- & -- \\
        & RepoLaunch   & 55.36 \textcolor{gray}{(62/112)} & 34 & 250 & 0.071 & 50.00 \textcolor{gray}{(25/50)} & 35 & 280 & 0.080 \\
        & SWE-Agent     & \underline{62.50} \textcolor{gray}{(70/112)}& 31& \textbf{200}& \textbf{0.058} & \underline{56.00} \textcolor{gray}{(28/50)} & 29 & \textbf{185} & \textbf{0.054} \\
        \rowcolor[gray]{0.95} &  \textbf{EnvPilot (Ours)} & \textbf{75.00} \textcolor{gray}{(84/112)}& \textbf{28} & \underline{220}& \underline{0.063}& \textbf{72.00} \textcolor{gray}{(36/50)} & \textbf{26} & \underline{198} & \underline{0.057} \\

    \bottomrule
  \end{tabular}
  }
    \begin{flushleft}
    \footnotesize \textbf{Note:} Bash-based Agent\cite{envbench} encountered technical issues with DeepSeek-V3-0324 as the model does not support the specific tool-calling framework. Repo2Run\cite{hu2025repo2runautomatedbuildingexecutable} is evaluated only on the Python subset, as its specialized toolset is specifically designed for Python environments and lacks compatibility with other programming languages.
    \end{flushleft}
    \vspace{-1em}
\end{table}

\textbf{Overall Performance.} \addedfinal{As can be seen from the overall results of the AES-Bench in Tab.~\ref{tab:cross_benchmark_comparison}, there are significant differences between the two foundation models and the three baseline agents and EnvPilot. Taking the Full dataset as an example, DeepSeek-V3-0324~\cite{liu2024deepseek} achieves success rates of 43.75\%, 55.36\%, 62.50\%, and 75.00\% on ExecutionAgent~\cite{executionagent}, RepoLaunch~\cite{swebenchgoeslive}, SWE-Agent~\cite{yang2024sweagent}, and EnvPilot, respectively. In contrast, GPT-4o-mini achieves only 23.21\%, 40.18\%, 24.11\%, and {46.43\%} respectively. Under the same task framework, the success rate of DeepSeek is generally 20\% to 40\% higher than that of GPT-4o-mini}. Even with the limitations of the smaller model, {EnvPilot significantly outperforms the baseline systems}. This result strongly validates the effectiveness of EnvPilot for environment setup tasks, particularly its core innovation component, the {TDM}. By retrieving and utilizing prior knowledge containing historical problems, solutions, and expert action sequences during the interaction, EnvPilot effectively avoids the blind trial-and-error observed in other baselines when facing unseen complex environment errors, thereby significantly improving the final task completion rate.

{\textbf{Resource Consumption.}} Regarding resource consumption, the token usage of EnvPilot does not increase significantly due to the introduction of TDM. On the contrary, it proves to be {more cost-efficient} in many scenarios. \addedfinal{With the DeepSeek-V3-0324, the token consumption of EnvPilot on the Full set is 220K, which is in the same order of magnitude as the 200K of SWE-Agent and clearly lower than the 420K of ExecutionAgent. While SWE-Agent maintains the lowest average cost (\$0.058), EnvPilot achieves a 12.50\% higher success rate with only a marginal increase of \$0.005.} This trend is more pronounced in GPT-4o-mini. While the small model causes large-scale token waste in baseline agents (e.g., the Full token consumption reaches 643K for ExecutionAgent, 387K for RepoLaunch, and 192K for SWE-Agent), EnvPilot uses {only 214K}, maintaining one of the lowest system-level costs. At the language level, the token cost of EnvPilot also remains within a controllable range. For instance, in JavaScript/TypeScript, RepoLaunch uses up to 560K tokens with GPT-4o-mini, while EnvPilot requires only 168K. This suggests that structured experience reduces invalid search paths, enabling the model to obtain more reliable experience-driven decisions with shorter reasoning chains.

{\textbf{Generalization.}} Cross-model generalization is another core observation of this experiment. Whether using DeepSeek-V3-0324 or GPT-4o-mini, {EnvPilot consistently outperforms the baseline systems} in success rate. \addedfinal{The generalization capability is further validated on ExecutionAgent-Bench, where EnvPilot achieves 72.00\% success, leading SWE-Agent (56.00\%) by a wide margin.} Under the DeepSeek-V3-0324, the improvement of EnvPilot over the baselines ranges between 10\% and 30\%. Under GPT-4o-mini, this advantage becomes even more apparent. The difference in token costs across the two models also shows a highly consistent trend, indicating that EnvPilot can achieve {stable transfer} between different models. This cross-model consistency profoundly reveals the value of TDM. When the capability of the backbone model is insufficient to solve environment conflicts independently, the TDM acts as {critical scaffolding}.

\textbf{Limitations of specialised workflows and scaling potential.} The experimental results further highlight the inherent limitations of highly specialised designs. At a basic level, non-agentic baselines such as Installamatic are completely unsuccessful (with a 0.00\% success rate) due to the lack of closed-loop execution feedback. Meanwhile, Repo2Run's over-reliance on Python-specific toolsets means it is incompatible with multi-language benchmarks.
Beyond these fundamental failures, a deeper cross-analysis of execution logs reveals a nuanced trade-off regarding framework paradigms and model scaling. Under the weaker GPT-4o-mini setup, the baseline agent RepoLaunch exhibits baseline stability on compilation-heavy languages (Java, C/C++). This is because RepoLaunch is a highly specialised workflow that relies on system-level engineering interventions (e.g.,  a ``Time Machine'' proxy) to bypass dependency version drift and artificially simplify the environment, thereby lowering the model's cognitive load. In contrast, EnvPilot relies on pure cognitive reasoning within a standardized baseline. When processing the dense error logs of Java and C/C++, the necessary multi-round interactions often exceed the cognitive capacity of GPT-4o-mini, resulting in failure due to instruction drift. However, upgrading the foundation model to DeepSeek-V3 eliminates this bottleneck entirely. With its enhanced long-context reasoning capabilities, EnvPilot can efficiently utilise the TDM to navigate complex dependencies, instantly overcoming its C/C++ deficit and achieving comprehensive dominance across all language subsets. This progression clearly shows that, although highly engineered, specialised workflows provide a stability baseline for weaker models, EnvPilot's general, experience-driven architecture has significant scaling potential and a much higher upper limit with stronger foundation models.

\addedfinal{\textbf{Failure Analysis and Performance Patterns.} Through a qualitative review of the execution traces, we identify three distinct performance characteristics of EnvPilot relative to the baselines.}
\addedfinal{First, efficiency in complex conflict resolution is a primary strength of EnvPilot. In scenarios involving missing system-level libraries or intricate build dependencies (e.g., \textit{dragonflydb\_\_dragonfly-5598}), baseline agents often struggle with redundant ``retry loops'' frequently wasting 5–10 interaction steps before failing. In contrast, EnvPilot utilizes the TDM to rapidly identify root causes—such as missing CMake configurations—and pivots its strategy immediately, significantly reducing the path to resolution.}
\addedfinal{Second, we observed a phenomenon of \textit{over-thinking in trivial tasks}. In a small subset of cases (e.g., \textit{pytest-dev\_\_pytest-5262}), a ``naive'' baseline agent may succeed by simply executing a direct command like \texttt{pip install}, whereas EnvPilot, guided by complex historical context, may attempt over-elaborate environment pre-checks or cleanup routines. These sophisticated preparations can occasionally introduce secondary issues, such as permission conflicts or timeouts, leading to failure where a simpler approach would have sufficed.}
\addedfinal{Finally, we note a \textit{performance convergence} in standard setups. In environments with minimal dependencies, the overhead of memory retrieval provides no significant advantage, and both EnvPilot and the baselines perform comparably. This suggests that, although EnvPilot is highly optimized for complex environmental setup tasks, it is primarily valuable for navigating non-trivial configuration conflicts that traditional agents cannot handle.}

\subsection{RQ2: Ablation Study on the Contribution of the TDM and the Context-aware Retrieval mechanism}
\label{sec:ablation_study}

To quantitatively evaluate the contributions of the core components of EnvPilot, we conducted a comprehensive ablation study. We analyzed how the TDM, the Context-aware Retrieval mechanism, and the structured memory format affect performance. We focused on both the success rate and the cost of execution. We compared EnvPilot against \addedfinal{five} ablated variants on the AES-Bench dataset. \addedfinal{Beyond the basic architectural components, we introduced two variants to compare our triplet structure with existing memory designs: Variant D (Problem + Solution) inspired by SWE-EXP~\cite{chen2025swe-exp}, and Variant E (Problem + Action) inspired by EvoConfig ~\cite{guo2026evoconfig}.} And we utilized DeepSeek-V3-0324, configuring the temperature to 0 to ensure determinism. Additionally, we retrieve the top-10 experiences to support the process.

\textbf{Variant Settings.} The full EnvPilot integrates TDM, the Context-aware Retrieval mechanism, and the execution pipeline. This configuration represents the performance upper bound. Variant A removes the TDM module entirely. It relies solely on the internal knowledge of the agent from pre-training and static codebase file analysis. Variant B retains TDM but replaces the Context-aware Retrieval mechanism with a naive query approach. Specifically, we embed \texttt{README.md} and \texttt{CI/CD} text into a query vector and search against the Problem field of the memory. Variant C substitutes the structured memory format with unstructured descriptions. The agent must retrieve relevant paragraphs from raw text. \addedfinal{To further analyze memory format, Variant D utilizes only the ``Problem'' and ``Solution'' fields (following SWE-EXP~\cite{chen2025swe-exp}) to test the impact of missing explicit action sequences. Variant E utilizes only ``Problem'' and ``Action'' (following EvoConfig~\cite{guo2026evoconfig}) to evaluate the necessity of abstract strategic guidance.}

\begin{table}[t]
  \centering
  \small
  \caption{Ablation results of TDM components and memory structures on AES-Bench using DeepSeek-V3-0324. Bold and underlined values represent the best and second-best results, respectively.}
  \label{tab:tdm_ablation}
  \resizebox{\textwidth}{!}{%
  \begin{tabular}{l|cccccccc|cc} 
    \toprule
    \multirow{2}{*}{\textbf{Variant}} & \multicolumn{8}{c|}{\textbf{ECSR (\%) $\uparrow$}} & \multicolumn{2}{c}{\textbf{Efficiency}} \\
    \cmidrule(lr){2-9} \cmidrule(lr){10-11}
    & \textbf{JS/TS} & \textbf{Java} & \textbf{Python} & \textbf{Go} & \textbf{Rust} & \textbf{PHP} & \textbf{C/C++} & \textbf{Total} & \textbf{API} $\downarrow$ & \textbf{Tokens} $\downarrow$ \\
    \midrule
    w/o TDM & 68.00 & 61.54 & 50.00 & 38.46 & 66.67 & \underline{75.00} & \textbf{68.00} & 62.50 & 31 & \textbf{200} \\
    w/ Naive Query & \underline{84.00} & \textbf{84.62} & 41.67 & \underline{69.23} & 75.00 & 58.33 & 52.00 & 66.96 & \textbf{28}& 219 \\
    w/ Unstructured & \textbf{88.00} & 69.23 & \textbf{66.67} & 61.54 & 66.67 & 66.67 & 48.00 & 66.96 & 29 & 217 \\
    w/ Prob + Solution & \underline{84.00} & \textbf{84.62} & \textbf{66.67} & 23.08 & \textbf{83.33} & 66.67 & 56.00 & 66.96 & \textbf{28}& \underline{204} \\
    w/ Prob + Action & 72.00 & 76.92 & 50.00 & \underline{69.23} & \textbf{83.33} & 66.67 & 56.00 & 66.96 & 29 & 206 \\
    \rowcolor[gray]{0.95} \textbf{w/ Full TDM} & 72.00 & \textbf{84.62} & \textbf{66.67} & \textbf{84.62} & \textbf{83.33} & \textbf{91.67} & \underline{60.00} & \textbf{75.00} & \textbf{28} & 220 \\
    \bottomrule
  \end{tabular}
  }
  \vspace{-1em}
\end{table}

\textbf{Analysis of Results.} Tab.~\ref{tab:tdm_ablation} presents the experimental results. \addedfinal{Overall, the full EnvPilot achieves the highest total success rate (75.00\%), outperforming all ablated variants (62.50\%--66.96\%). This result supports our main claim: combining the complete \texttt{<problem, solution, action>} experience format with the Context-aware Retrieval mechanism is necessary to fully unlock the benefit of experience reuse.} \addedfinal{Compared to the strongest ablated setting (66.96\%), the full system yields a +8.04 percentage-point gain in success rate.}

\addedfinal{Variant A (w/o TDM) causes the largest performance degradation. The total success rate drops from 75.00\% to 62.50\% (-12.50 points), indicating that TDM is the most critical component for robust environment setup across diverse repositories. Without external memory, the agent can only rely on the backbone model and static codebase inspection, which is often insufficient for resolving non-trivial dependency conflicts, toolchain mismatches, or rare build errors.} \addedfinal{Interestingly, Variant A shows the highest API usage (31) but the lowest token consumption (200K), suggesting that failing cases may terminate earlier with shorter traces, even though the agent makes more trial-and-error calls.}

\addedfinal{Variants B--E (w/ Naive Query, w/ Unstructured, w/ Problem+Solution, w/ Problem+Action) all reach the same total success rate (66.96\%). This indicates that simply introducing memory is not enough: without accurate context grounding (Variant B), without an explicit structured schema (Variant C), or with incomplete triplet fields (Variants D/E), the agent cannot consistently transform retrieved experience into correct, executable decisions.} \addedfinal{In other words, our gains mainly come from the \emph{complete} triplet structure plus \emph{context-aware} retrieval, rather than from memory insertion alone.}

\addedfinal{At the per-language level, the full EnvPilot provides the most consistent improvements in Java (84.62\%), Go (84.62\%), Python (66.67\%), Rust (83.33\%), and PHP (91.67\%), where setup outcomes are highly sensitive to build systems and dependency resolution. In particular, the gains over the best ablated variants are substantial for Go (+15.39 points) and PHP (+16.67 points).} \addedfinal{For Rust, the full system ties with the best ablations (83.33\%), suggesting that partial fields can already capture most useful signals for these tasks.} \addedfinal{We also observe two notable exceptions: (i) for JavaScript, unstructured memory performs best (88.00\%), implying that free-form textual cues may already be sufficient for many JavaScript setups; (ii) for C/C++, removing TDM achieves the highest success rate (68.00\%), while memory-based variants are lower (48.00\%--60.00\%), indicating that heterogeneous toolchains and subtle context differences can make retrieved experience easier to misapply in C/C++ projects. These results motivate future improvements on context inference and negative filtering for toolchain-sensitive languages.}

\addedfinal{Regarding efficiency, the full EnvPilot achieves the lowest API calls (28), showing that better retrieval and decision quality reduce unnecessary iterations. However, the token usage of the full system (220K) is not the smallest, because richer context-aware retrieval and structured experience injection increase the per-step prompt length. Overall, the results demonstrate a clear accuracy--cost trade-off: EnvPilot attains markedly higher success with fewer API calls, at the price of a modest token overhead.}

\subsection{\addedfinal{RQ3: Reliability of Automated Evaluation}}
\label{sec:eval_validation}
\addedfinal{To validate the reliability of our automated evaluation, we conducted a rigorous large-scale human verification. We focused on assessing the agreement between our LLM-as-a-Judge system and human experts.}

\addedfinal{\textbf{Experimental Setup.}
We constructed a validation dataset consisting of 460 execution trajectories. This set encompasses the complete collection of experimental outputs generated by DeepSeek-V3-0324 on AES-Bench in RQ1, serving as the comprehensive subject for our manual annotation. Human experts reviewed these trajectories to establish the ground truth labels (Success or Failure) based on the criteria defined in Section~\ref{sec:hybrid_eval}.}

\addedfinal{\textbf{Results.}
We compared the decisions made by the LLM judge against the human ground truth. As shown in Tab.~\ref{tab:eval}, the LLM judge demonstrates exceptional reliability. Out of 460 instances, the system correctly identified 267 successes (True Positives) and 184 failures (True Negatives). Discrepancies were minimal, with only 2 False Positives and 7 False Negatives. Quantitatively, the evaluation system achieved an Accuracy of 98.04\% and an F1 Score of 98.34\%. The high Precision (99.26\%) indicates that the model is extremely conservative in declaring success, effectively minimizing false positives. Meanwhile, the Recall of 97.45\% confirms that it captures the vast majority of valid configurations. These results confirm that the pure LLM-as-a-Judge approach is highly consistent with human judgment and serves as a reliable standard for evaluating automated environment setup tasks.}

\begin{table}[h]
  \centering
  \small
  \caption{\addedfinal{Performance of LLM-as-a-Judge against Human Validation (N=460).}}
  \label{tab:eval}
  \setlength{\tabcolsep}{10pt} 
  \begin{tabular}{ccccc}
    \toprule
    \textbf{Total Samples} & \textbf{Accuracy} & \textbf{Precision} & \textbf{Recall} & \textbf{F1 Score} \\
    \midrule
    460 & 98.04\% & 99.26\% & 97.45\% & 98.34\% \\
    \bottomrule
  \end{tabular}
\end{table}

\subsection{RQ4: Impact of Experiences Number}

\addedfinal{To study how the number of retrieved experiences affects the agent's behavior, we vary the size of the experience memory and evaluate the agent under five settings: 0, 5, 10, 15 and 20 retrieved experiences.} We use DeepSeek-V3-0324 as the backbone model, EnvPilot serves as the agent for executing automated environment setup tasks. All experiments are conducted on AES-Bench. For each setup task, we record (1) the success rate, (2) the avg. API calls, and (3) the avg. token consumption. These metrics provide a comprehensive view of the trade-offs between effectiveness and efficiency.

\begin{figure*}[ht]
  \centering
  \includegraphics[width=\textwidth]{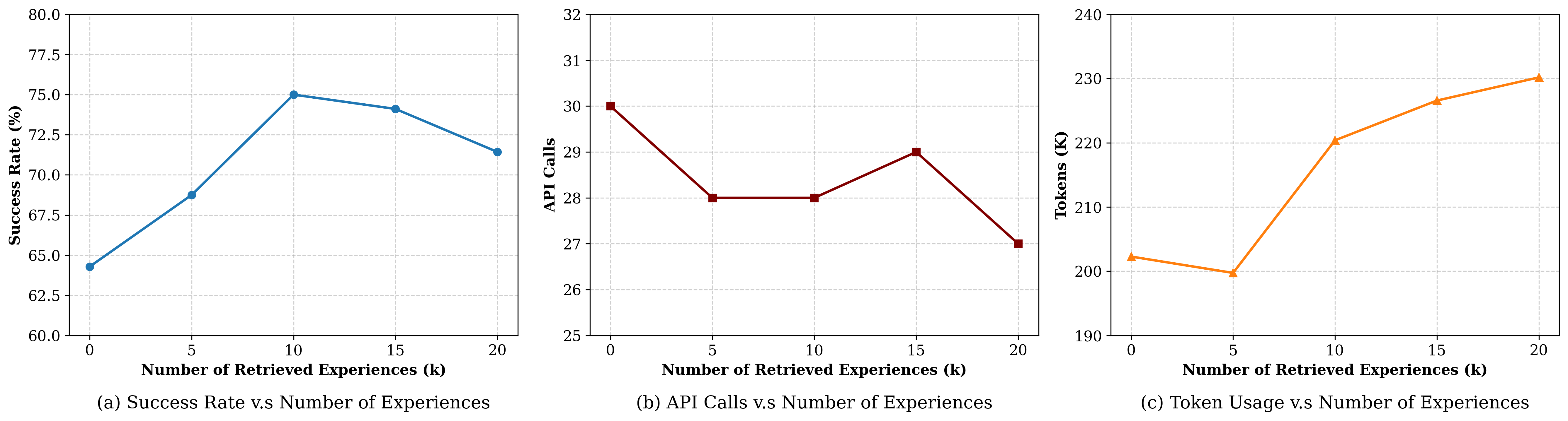}
  \caption{Trends on the Full dataset across different retrieval experiences. Subfigure~(a) shows success rate, (b) shows API calls, and (c) shows token usage. The results indicate that $k=10$ offers the most balanced performance.}
  \label{fig:trends_with_topk}
    \vspace{0em}
\end{figure*}

\textbf{Overall Trends.} \addedfinal{Fig.~\ref{fig:trends_with_topk} demonstrates the effect of the number of retrieved experiences (set to 0, 5, 10, 15 and 20) on success rate, API calls, and token usage across different programming languages. Overall, the results show that the number of retrieved experiences has a non-linear impact on performance. The average success rate significantly increases from 68.75\% to 75.00\% as $k$ grows from 5 to 10. However, a further increase to $k=15$ and $k=20$ leads to a performance decline, with the success rate dropping to 74.11\% and 71.43\%, respectively. 
This suggests that a moderate amount of context effectively helps the model understand the task. Conversely, excessive information (e.g., $k=20$) can lead to attention dilution and contextual noise. This makes it difficult for the model to identify the most relevant experience among many similar ones.} This phenomenon of information overload negates the benefits of additional context, indicating that merely increasing the retrieval quantity does not guarantee superior task outcomes.

\begin{table}[ht]
  \centering
  \small
  \caption{Performance and cost trends across different numbers of retrieved experiences ($k$). Bold and underlined values represent the best and second-best results, respectively.}
  \label{tab:retrieved_experience_transposed}
  \resizebox{\textwidth}{!}{%
  \begin{tabular}{l|cccccccc|cc} 
    \toprule
    \multirow{2}{*}{\textbf{\# Exp. ($k$)}} & \multicolumn{8}{c|}{\textbf{ECSR (\%)}} & \multicolumn{2}{c}{\textbf{Efficiency}} \\
    \cmidrule(lr){2-9} \cmidrule(lr){10-11}
    & \textbf{Java} & \textbf{JS/TS} & \textbf{C/C++} & \textbf{Rust} & \textbf{Go} & \textbf{Python} & \textbf{PHP} & \textbf{Total} & \textbf{API} $\downarrow$ & \textbf{Tokens(K)} $\downarrow$ \\
    \midrule
    $k=0$  & 61.54 & \underline{76.00}& \underline{64.00} & 58.33 & 53.85 & \underline{58.33} & 66.67 & 64.29 & 30 & \underline{202} \\
    $k=5$  & 69.23 & \underline{76.00}& 56.00 & \textbf{91.67} & 69.23 & 50.00 & \underline{75.00} & 68.75 & \underline{28} & \textbf{200} \\
    \rowcolor[gray]{0.95} $k=10$ & \textbf{84.62}& 72.00& 60.00& \underline{83.33} & \textbf{84.62} & \textbf{66.67}& \textbf{91.67} & \textbf{75.00}& \underline{28}& 220\\
    $k=15$ & \underline{76.92} & \textbf{80.00}& \textbf{80.00} & 75.00 & 69.23 & 50.00 & 75.00 & \underline{74.11} & 29 & 227 \\
    $k=20$ & \underline{76.92} & 72.00 & \underline{64.00} & 75.00 & \underline{76.92} & \underline{58.33} & 83.33 & 71.43 & \textbf{27} & 230 \\
    \bottomrule
  \end{tabular}
  }
\end{table}

\textbf{Resource Consumption.} \addedfinal{Regarding cost, the token usage increases gradually as more experiences are retrieved, rising from 200K tokens ($k=5$) to 230K tokens ($k=20$). The growth is expected, since larger retrieval sets introduce more content into the model’s context. However, the performance gain does not scale proportionally with the cost increase. The $k=10$ configuration stands out as the most cost-effective choice. It provides a 6.25\% improvement in success rate compared to $k=5$. The average API calls remain stable across all settings, fluctuating between 27 and 30 calls. This confirms that while retrieval depth increases the context size per prompt, it has minimal impact on the number of logical steps required by the agent to complete the task.}

\textbf{Language Level.} \addedfinal{Tab.~\ref{tab:retrieved_experience_transposed} details the impact of retrieved experience quantity ($k$) across different languages. The results confirm that $k=10$ represents the optimal trade-off between context and noise.
First, for standardized ecosystems like Java, Go, and PHP, the agent achieves peak or near-peak performance at $k=10$, with success rates reaching 84.62\%, 84.62\%, and 91.67\% respectively. This indicates that 10 experiences are sufficient to cover most common setup patterns in these languages.
Second, increasing the number of experiences beyond 10 generally leads to diminishing returns. As $k$ rises to 20, the average success rate drops to 71.43\%. Although complex languages like C/C++ show slight fluctuations due to toolchain heterogeneity, the overall trend clearly shows that excessive retrieval ($k=20$) introduces noise that outweighs the benefits of additional context. Therefore, we select $k=10$ as the default setting for EnvPilot.}

In summary, retrieving 10 experiences provides the best balance between effectiveness and computational cost. It offers enough contextual signals to guide the agent while avoiding the noise introduced by larger retrieval sizes. Larger values such as $k=15$ and $k=20$ lead to diminishing returns and even degrade performance, \addedfinal{confirming that the quality and relevance of retrieved experiences are far more critical than sheer quantity.}

\subsection{\addedfinal{RQ5: Impact of LLM Nondeterminism and Result Stability}}
\label{sec:stability_analysis}

\addedfinal{To mitigate the inherent non-determinism of Large Language Models (LLMs) and ensure the robustness of our findings, we conducted three independent trials for both the full EnvPilot (w/ TDM) and the ablated version (w/o TDM) on the AES-Bench dataset. We utilize the DeepSeek-V3-0324 as the generative model, and set temperature to 0. While the temperature was set to 0 to prioritize determinism, minor fluctuations in model outputs and external environment factors (e.g., network latency or API response variations) can still influence the final success rate.}

\begin{table}[htbp]
  \small
  \caption{Stability analysis of w/o TDM vs. w/ TDM.}
  \label{tab:stability_results}
  \begin{tabular}{llcccc}
    \toprule
    \textbf{Variant} & \textbf{Trial} & \textbf{ECSR (\%) $\uparrow$} & \textbf{API $\downarrow$} & \textbf{Token (K) $\downarrow$} & \textbf{Cost (\$) $\downarrow$} \\
    \midrule
    \multirow{4}{*}{w/o TDM} 
      & Trial 1 & 64.29 \textcolor{gray}{(72/112)} & 30 & 202 & 0.058 \\
      & Trial 2 & 62.50 \textcolor{gray}{(70/112)} & 31 & 200 & 0.058 \\
      & Trial 3 & 59.82 \textcolor{gray}{(67/112)} & 30 & 194 & 0.056 \\
      \cmidrule{2-6}
      \rowcolor[gray]{.95} &  \textit{Mean} & 62.20 $\pm$ 2.25 & 30 $\pm$ 0.58 & \textbf{199 $\pm$ 4.16} & \textbf{0.057 $\pm$ 0.001} \\
    \midrule
    \multirow{4}{*}{{w/ TDM}} 

      & Trial 1 & 75.00 \textcolor{gray}{(84/112)} & 28 & 220 & 0.063 \\
      & Trial 2 & 69.64 \textcolor{gray}{(78/112)} & 28 & 221 & 0.063 \\
      & Trial 3 & 77.68 \textcolor{gray}{(87/112)} & 27 & 212 & 0.061 \\
      \cmidrule{2-6}
      \rowcolor[gray]{.95} & {\textit{Mean}} & \textbf{74.11 $\pm$ 4.09} & \textbf{28 $\pm$ 0.58} & 218 $\pm$ 4.93 & 0.062 $\pm$ 0.001 \\
    \bottomrule
  \end{tabular}
\end{table}

\noindent \addedfinal{\textbf{Stability Analysis.} Tab.~\ref{tab:stability_results} reports three independent trials for both EnvPilot and the ablated baseline without TDM. EnvPilot achieves a mean success rate of 74.11\% $\pm$ 4.09, while the w/o TDM baseline reaches 62.20\% $\pm$ 2.25. This corresponds to an absolute improvement of +11.91 percentage points, i.e., a +19.14\% relative gain over the baseline.}
\addedfinal{The advantage is consistent across trials. The lowest success rate of EnvPilot (69.64\%) is still higher than the best-case outcome of the w/o TDM baseline (64.29\%). This indicates that TDM raises a reliable performance floor, and also increases the performance ceiling (up to 77.68\% in Trial 3).}
\addedfinal{In terms of execution overhead, EnvPilot requires fewer API calls on average (28 $\pm$ 0.58 vs. 30 $\pm$ 0.58), suggesting that structured experience retrieval reduces unnecessary trial-and-error steps. However, EnvPilot consumes more tokens (218K $\pm$ 4.93 vs. 199K $\pm$ 4.16), and thus incurs a slightly higher average cost per repository (0.062 $\pm$ 0.001 vs. 0.057 $\pm$ 0.001). This increase is expected because the input length is expanded by injecting retrieved experiences and context-aware prompts, even when fewer calls are made. Overall, the results confirm that the improvement is systematic rather than random. This delivers a clear gain in accuracy with stable and bounded overhead.}

\subsection{RQ6: Case Study and Behavioral Interpretation}
\label{sec:case_study}
\begin{figure}[ht]
    \centering
    \includegraphics[width=\textwidth]{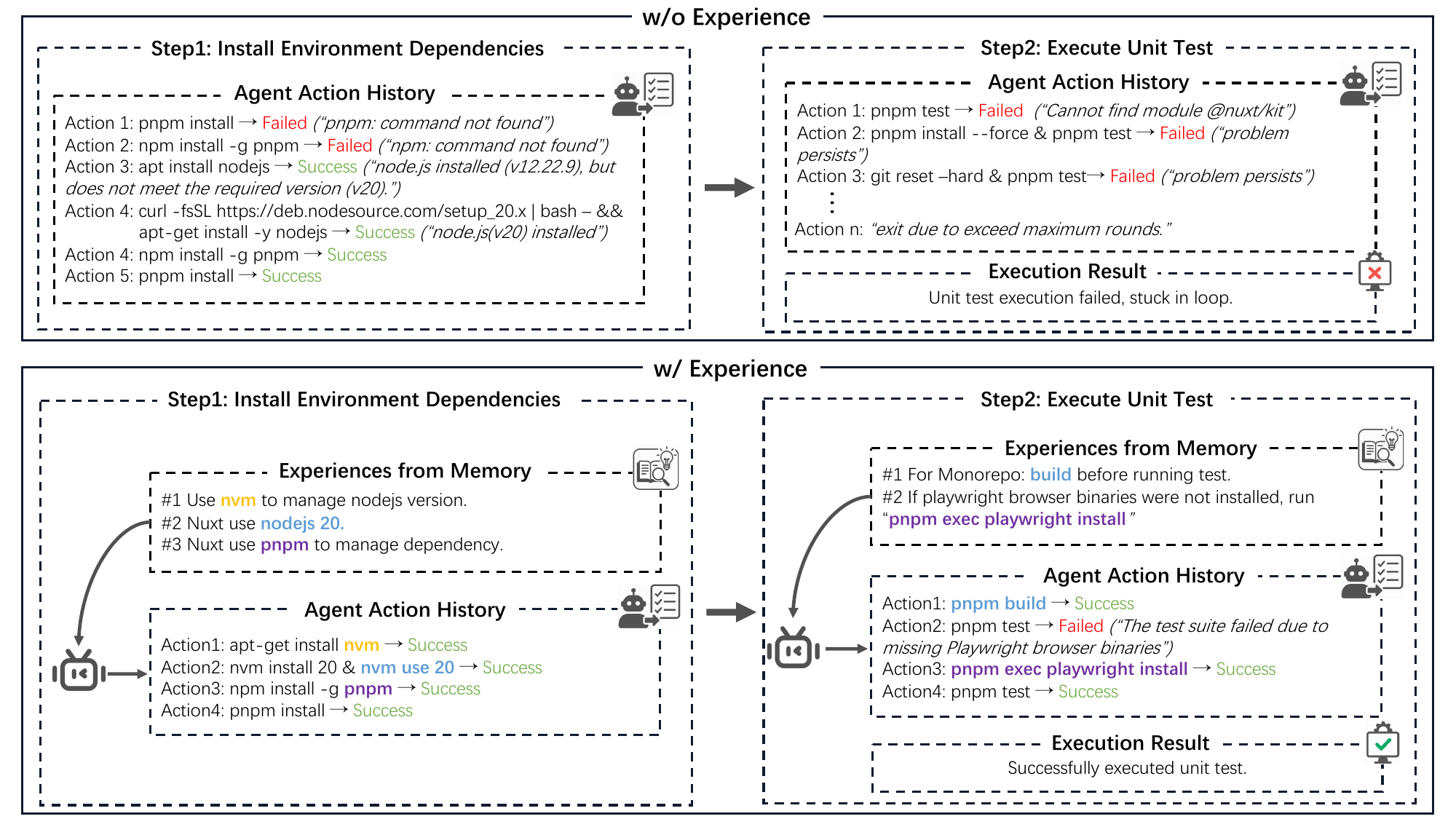}
    \caption{The impact of our experience mechanism on task execution. 
    The agent without experience (top) fails due to inefficient trial-and-error. In contrast, the agent with experience (bottom) leverages stored knowledge to follow a direct and correct action sequence, successfully completing the task.
    }
    \vspace{0em}
    \label{fig:case_study}
\end{figure}

To provide an intuitive demonstration of the proposed TDM, a case study was presented in Fig.~\ref{fig:case_study}. It compared the agent's performance with and without access to TDM when tasked with configuring a \texttt{Nuxt.js} project environment and executing unit tests in a controlled setting.

\textbf{Without Experience: A Trial-and-Error Struggle.}
As illustrated in the upper panel, the agent without prior experience followed a lengthy and ineffective trial-and-error process.
In Step 1 (Install Environment Dependencies), the agent first attempted to install dependencies with \texttt{pnpm install} and \texttt{npm install}, but found that Node.js was not installed.
It then used the \texttt{apt} tool to install Node.js, which resulted in an outdated version (\texttt{v12.22.9}). This indicated that the agent was not familiar with the version management tools commonly used in Node.js projects.
In Step 2 (Execute Unit Test), the agent was unable to run the tests due to a missing module and attempted to resolve the issue using generic commands, such as \texttt{``pnpm install --force''} and \texttt{``git reset --hard''}.
These measures failed to resolve the root problem, which lay in an unexecuted build step required by the monorepo structure.
As a result, the agent became stuck in a repetitive cycle and ultimately failed to complete the task.

\textbf{With Experience: A Guided and Efficient Path.}
By contrast, the lower panel depicted an agent equipped with experience memory. In Step 1, it recalled clear key instructions: ``Use nvm to manage Node.js versions'', ``Nuxt requires Node.js 20'', and ``Nuxt uses pnpm''. Guided by this knowledge, the agent followed a concise and correct sequence. It first installed \texttt{nvm}, then set the correct Node.js version, and finally installed the project dependencies with \texttt{pnpm}.
In Step 2, it applied additional prior knowledge, such as ``For Monorepo: build before running tests'' and guidance on resolving missing Playwright binaries. Accordingly, it executed \texttt{``pnpm build''} first. When it encountered a failure due to missing browser binaries, it applied the precise and previously validated solution from memory (\texttt{``pnpm exec playwright install''}), which enabled the successful execution of unit tests.

\textbf{Mechanism.} This case study shows that TDM can elevate an agent from a naive problem solver to a context-aware assistant. By recalling salient precedents from prior trajectories, the agent prunes unproductive branches early, suppresses repetitive trial-and-error loops, and delivers guidance conditioned on the project’s toolchain, coding conventions, and deployment constraints. As a result, the agent performs fewer redundant actions and repeats fewer mistakes. Its advice also matches local requirements more closely, which improves execution efficiency and increases the overall task success rate in realistic settings.

\textbf{Test-time Scaling.} Viewed through the lens of test-time scaling, TDM offers a pathway to continual learning. Without altering base-model parameters, the system accumulates and curates high-value trajectories and, through their selective reuse, steadily improves decision quality with use. This constitutes a genuine form of continual learning. Post-inference, knowledge is consolidated into retrievable procedural units and invoked immediately when similar situations recur. For example, ``build before test'', ``remedies for missing runtime binaries'', or ``minimal-change resolutions for version conflicts''. As the TDM scales, the system’s effective prior experience strengthens, exhibiting a “the more it is used, the more accurate it becomes” scalability.

\subsection{RQ7: Failure Mode Analysis}
\begin{figure}[tbp]
    \centering
    \includegraphics[width=0.72\linewidth]{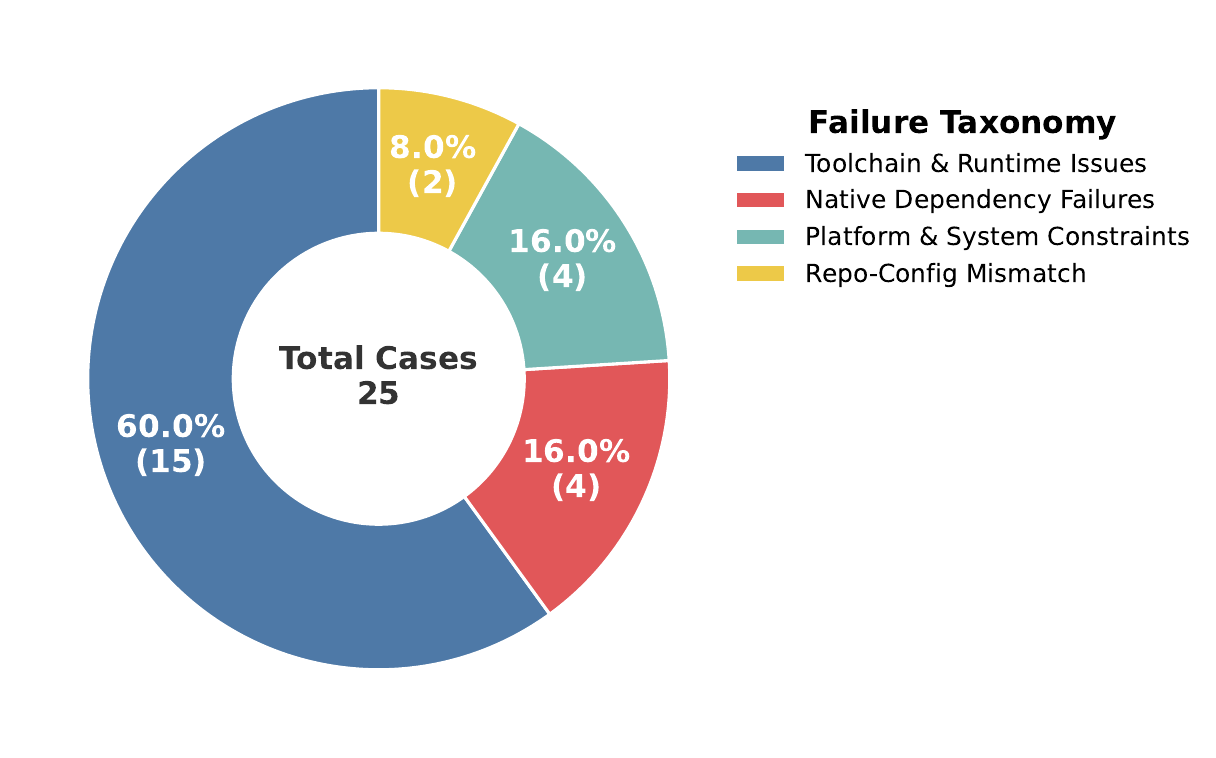}
    \caption{Distribution of Failure Causes in EnvPilot}
    \label{fig:failure_taxonomy}
\end{figure}

\addedfinal{
We analyze the failures from the best-performing trial (Trial 3, 77.68\%) in RQ5 to characterize the residual failure modes that persist even under the best observed outcome.
While EnvPilot achieves an outstanding success rate, analyzing the remaining 25 failed instances provides critical insights into the boundaries of automated environment setup. We conducted a manual inspection of the execution logs produced by the DeepSeek-V3-0324 model. We categorized the root causes into four primary taxonomies, with the distribution detailed in Fig.~\ref{fig:failure_taxonomy}.}

\addedfinal{
The most dominant failure category is Toolchain and Runtime Issues (60\%). This category includes version incompatibilities and environment activation failures. The agent often identifies the correct tool but fails to satisfy specific version constraints or manage shell states. For instance, in the \textit{astral-sh/uv} project, the agent attempted to use an older Rust compiler (version 1.75). However, the project required a newer edition that was not supported by the installed toolchain. Another common issue involves environment activation. In the \textit{nuxt/nuxt} project, the agent successfully installed the \textit{pnpm} package manager. However, it failed to execute the source command to update the \textit{PATH variable}. As a result, subsequent commands could not locate the installed tool.
The second major challenge involves system-level constraints and native dependencies. Platform and System Constraints (16\%) arise when the containerized environment restricts necessary privileges or services. For example, the \textit{nasa/openmct} project failed because the container blocked \textit{ChromeHeadless} from running without the sandbox flag. Additionally, Native Dependency Failures (16\%) occur during the compilation of C++ or Python extensions. The \textit{scikit-learn} project failed due to mismatches in the BLAS toolchain, and the \textit{mpv-player} project encountered ABI incompatibility with the system \textit{libavcodec} library. These failures indicate that configuring complex environments requires precise alignment of system-level binary interfaces beyond simple package installation. Finally, Repo-Config Mismatch (8\%) happens when the agent misclassifies the project type. For instance, the agent incorrectly identified the C-based \textit{php/php-src} project as a Go repository.}

\addedfinal{
Overall, most failures come from toolchain and runtime issues.
They account for 60\% of the failed cases.
This shows that correct packages are not enough.
The agent must also satisfy version constraints and keep a valid shell state.
The remaining failures are split between platform constraints and native dependency issues, each at 16\%.
They highlight the need for better handling of container restrictions and system level binary compatibility.
Although repo config mismatch is less frequent at 8\%, it can still lead to incorrect setup plans. These findings suggest that robust version control, stable environment activation and system dependency alignment are essential for further improving EnvPilot.}

\section{Related Work}
\addedfinal{
\textbf{Software Engineering Configuration, Dependencies, and Reproducibility.}
(1) \emph{Software Configuration Management and Configuration Management.}
Automated environment setup is closely related to long-standing software configuration management (SCM) and configuration management (CM) research in software engineering.
Classic SCM literature emphasizes controlling and evolving complex software artifacts, including workspace control and build/release management as core concerns, and highlights that maintaining consistent configurations over time is practically difficult due to frequent changes across tools, dependencies, and platforms~\cite{Estublier2000SCMRoadmap}.
In addition, CM systems and ``desired-state'' configuration tools (e.g., policy- or rule-based system configuration) can automate deployments when the target configuration is explicitly specified.
However, such systems typically assume that configuration specifications (e.g., manifests/recipes) are available and correct, and provide limited support when specifications are incomplete or when failures require iterative debugging guided by build/test logs.}
\addedfinal{
(2) \emph{Dependency and Build Management.}
Language package managers and build tools (e.g., Maven/Gradle, npm/yarn, pip/poetry, cargo) automate dependency resolution and compilation/testing given declared metadata.
Yet empirical SE studies show that developers often delay dependency updates and that dependency-related issues (including conflicts and breakages) are prevalent and non-trivial to diagnose and fix in practice~\cite{Kula2018DepUpdateESE}.
These observations motivate automation beyond ``install what is declared'': real-world environment setup frequently requires interpreting failure signals, inferring missing/implicit dependencies or tooling requirements, and iteratively repairing scripts and commands, especially when repositories span multiple languages and heterogeneous build conventions.}
\addedfinal{
(3) \emph{Reproducible and Hermetic Builds.}
Reproducible builds aim to ensure identical build outputs given the same source and dependencies, enabling independent verification and improving supply-chain integrity~\cite{LambZacchiroli2022ReproducibleBuilds}.
Practical approaches include containerization and functional package management, such as Docker and Nix/Guix-style systems, which improve isolation and offer declarative capture of dependencies~\cite{Merkel2014Docker,Dolstra2004Nix,Courtes2013Guix}.
Nevertheless, these approaches still rely on correct and complete environment specifications (e.g., Dockerfiles, pinned inputs, build rules) and do not eliminate the need for iterative troubleshooting when upstream build scripts, transitive dependencies, or external tooling assumptions change.}

\textbf{LLMs for Software Engineering.}
The rapid development of LLMs is profoundly transforming the field of software engineering (SE). Prior studies have demonstrated that LLMs significantly improve developer productivity and efficiency, positioning them as essential tools in modern SE workflows. For example, intelligent programming assistants such as GitHub Copilot have been shown to reduce development time considerably, highlighting the disruptive potential of LLMs in programming practices. These advances suggest that LLMs are evolving beyond auxiliary tools and are driving software engineering toward higher levels of automation and intelligence~\cite{swebench, mswebench, swebenchgoeslive, swegym}.
In terms of applications, LLMs have been explored in multiple directions. Code generation is one of the most established use cases, where models can produce executable code directly from natural language descriptions and achieve near-human performance in solving complex programming 
tasks~\cite{chen2021evalllmoncode, li2022competitionlevelcode, jiang2024survey, yang2024less, joel2410survey}. 
Code completion has also been studied widely. Many studies indicate that code completion is moving toward context-aware and user-aware assistants~\cite{tan2024prompt, giagnorio2025personalizing}.
Meanwhile, bug fixing has emerged as a critical research focus, with models capable of automatically locating and repairing defects based on issue descriptions and error messages, thereby simulating the debugging process of experienced engineers ~\cite{swebench, mswebench, swebenchgoeslive, swerebench}. 

Furthermore, LLMs have shown potential in code review, where they can identify potential issues in code changes and provide constructive suggestions to improve software quality ~\cite{lu2023llamacodereview, sun2025autocodereview}. In addition, LLMs have been applied to automated testing, generating test cases and assertions that enhance coverage and accelerate defect detection ~\cite{zhang2023llmgentestwell, chen2024chatunitest, ryan2024codeawaretestgen}. Collectively, these applications demonstrate the wide-ranging prospects of LLMs across different stages of the software engineering lifecycle and highlight their growing practical value.

\textbf{LLM-based Agents in Environment Setup.}
Early methods such as Installamatic~\cite{autoinstallinpy} explored inferring installation steps directly from project documentation, taking the first steps toward automated environment setup. However, its reliance on documentation quality made it prone to failure when information was incomplete or dependencies were complex, limiting its ability to handle diverse real-world projects.
ExecutionAgent~\cite{executionagent} introduced meta-prompt and execution-based feedback, which significantly improved cross-language environment setup and test execution. However, it remained dependent on trial-and-error, making it vulnerable to unstable outputs and execution failures, particularly when dealing with complex build tools or nonstandard environments, where robustness was limited. 
Repo2Run~\cite{hu2025repo2runautomatedbuildingexecutable} proposed a complete agent architecture that integrates ``atomic configuration synthesis'', dual-environment isolation with rollback, and automatic Dockerfile generation. This design greatly enhanced controllability and traceability. Still, it was restricted to Python repositories and lacked strong support for complex dependencies, advanced custom scripts, or nonstandard base images. \addedfinal{Recently, EvoConfig\cite{guo2026evoconfig} introduced a self-evolving mechanism to dynamically adjust diagnostic rules during the configuration process. However, EvoConfig focuses on intra-task adaptation and explicitly states that it does not rely on external memory modules, limiting its ability to transfer knowledge across different repositories. In contrast, EnvPilot leverages a cross-task Trajectory-Derived Memory, enabling the agent to retrieve and reuse validated solutions from historical tasks to solve novel problems more efficiently.}

\addedfinal{\textbf{Memory Systems in LLM-based Agents.}
Memory has emerged as a core mechanism for enabling experience reuse in LLM-based agents, particularly in execution-centric software engineering (SWE) tasks.
Rather than serving solely as a cognitive aid for long-horizon reasoning, recent work increasingly treats memory as a structured repository of actionable experience, capturing concrete execution outcomes such as configuration decisions, failure patterns, and validated actions. Existing memory systems can be broadly analyzed along three design dimensions: memory representation, experience integration, and retrieval strategies. (1) \emph{Memory representation.} Early approaches, such as MemoryLLM~\cite{wang2024memoryllm}, embed knowledge into latent representations, offering flexibility but limited interpretability and weak grounding in execution artifacts.
Subsequent studies have explored more explicit and structured memory formulations to improve controllability and transferability, including graph-based organizations~\cite{anokhin2024arigraph}, as well as hierarchical architectures that separate short-term context from long-term storage~\cite{packer2023memgpt}.
More recent work extends structured memory to multimodal settings~\cite{li2024optimus} or enforces modular reasoning through explicit state abstractions such as finite-state machines~\cite{guan2024amor}.
While these approaches improve organization and reasoning stability, they primarily store abstracted summaries or state-level information, which are not directly executable in SWE workflows. 
(2) \emph{Experience integration.} A complementary line of research focuses on how experiences are distilled, integrated, and reused across tasks.
KnowAgent~\cite{zhu2024knowagent} integrates external action--knowledge bases into agent prompts.
AgentKB~\cite{tang2025agent} further proposes a shared cross-task knowledge base with teacher--student abstraction, enabling systematic refinement and transfer of experience across domains.
Formal-LLM~\cite{li2024formal} and AutoAct~\cite{qiao2024autoact} introduce explicit structural constraints, such as formal rules or constructed action spaces, to stabilize reasoning and facilitate transferable behavior.
More recently, SWE-Exp~\cite{chen2025swe-exp} highlights the importance of experience-driven memory in software engineering by distilling reusable knowledge from prior agent trajectories, including both successful and failed executions.
By organizing experiences at multiple granularities and grounding them in concrete development actions, SWE-Exp demonstrates that continuous experience accumulation is critical for robust performance in SWE tasks. In a similar spirit, recent work on experience replay and record--replay mechanisms~\cite{feng2025get, liu2025contextual} emphasizes capturing and reusing validated execution traces to improve agent robustness and efficiency.
These approaches demonstrate that continuous accumulation and reuse of execution experience is critical for reliable agent behavior.
Despite these advances, most existing methods remain reasoning-centric and do not explicitly target
execution-level experience reuse for environment setup and dependency resolution. (3) \emph{Retrieval strategies.}
Beyond memory construction, retrieval mechanisms critically influence how effectively stored experiences can be reused.
Several approaches extend standard retrieval-augmented generation (RAG)~\cite{lewis2020retrieval} by incorporating biologically inspired indexing~\cite{jimenez2024hipporag}, temporal cues~\cite{liu2025echo}, hierarchical subgoal structures~\cite{hu2024hiagent}, or state-aware guidance conditioned on environmental context~\cite{fu2024autoguide}.
These techniques improve retrieval relevance in long-horizon reasoning tasks, but typically assume static or externally specified queries, limiting their applicability to dynamic, project-specific SWE settings.
}

\addedfinal{More recently, experience replay and record--replay mechanisms have explored how to capture and reuse validated execution traces to improve agent robustness and efficiency.
Contextual Experience Replay (CER)~\cite{liu-etal-2025-contextual} accumulates past episodes and replays selected experiences by converting them into natural-language experience descriptions injected into the context window, which is effective for sequential decision-making settings (e.g., web navigation) where contextual hints can be directly reused.
AgentRR~\cite{feng-etal-2025-get} proposes a record--summary--replay framework that summarizes recorded executions at multiple abstraction levels and replays them under a trusted checking function, aiming to improve reliability, privacy, and cost for general agent execution.
However, these paradigms are not a drop-in solution for automated software environment setup.
First, setup traces and logs are often long and noisy, making in-context replay token-inefficient and hard to map from symptoms to actionable fixes.
Second, environment setup requires \emph{composable and transferable} remediation units across highly heterogeneous toolchains (languages, package managers, build systems), whereas replaying full workflows frequently overfits to a specific repository and generalizes poorly when dependency stacks or commands differ.
Third, many failures can be anticipated from repository artifacts (README/CI/build scripts/config) before execution, calling for proactive, context-aware retrieval rather than purely reactive replay after failures.
Motivated by these task characteristics, EnvPilot distills trajectories into structured, executable $\langle$problem, solution, action$\rangle$ experiences and performs proactive retrieval grounded in codebase profiling with hypothetical query generation and multi-query retrieval/reranking.}

\textbf{Benchmarks in Environment Setup.}
Early benchmarks, such as Installamatic~\cite{autoinstallinpy}, provided an initial evaluation standard for environment setup in Python repositories. Covering around 40 projects, it helped demonstrate the basic feasibility of agent-based approaches, but its limited scale made it insufficient to capture the diversity of real-world scenarios. 
ExecutionAgent~\cite{executionagent} proposed a benchmark centered on test execution. For each instance, it supplied deployment and testing scripts to measure whether an agent could successfully build the project, run its tests, and produce results consistent with the ground truth through manual verification. While this method more closely reflected real usage, it remained constrained by the small number of projects and the human effort required for evaluation. 
EnvBench~\cite{envbench} established an automated evaluation framework. It included 329 Python and 665 JVM (Java/Kotlin) repositories, with a stronger focus on projects that are difficult to set up automatically. EnvBench introduced automated evaluation metrics: static analysis to detect missing dependencies in Python projects, and compilation checks for JVM projects. This lightweight, automated approach significantly improved scalability and reproducibility. However, it still covered only Python and Java, leaving the broader diversity of real-world software engineering scenarios insufficiently represented.
To address these limitations, we propose AES-Bench, which incorporates a broader range of environment setup tasks and introduces an automated evaluation framework for more efficient and comprehensive assessment.

\section{Threats to Validity}
We discuss three types of validity and how we address them.

\textbf{Threats to external validity.}
External validity concerns how well our results generalize to other settings. We design AES-Bench to improve this aspect. It contains 112 open source instances from 9 languages and different ecosystems. The projects use different build tools and package managers, such as pip, npm, yarn and others. This variety helps us cover a wide range of environment setup patterns. We also choose strong and widely used open source LLMs as the base models. They support tool use and long contexts, which matches the needs of agent based environment setup. These choices make our findings more relevant to current practice.
However, several threats to external validity remain. Our benchmark focuses on open source repositories with public dependencies and standard registries. Industrial projects may rely on private mirrors, internal scripts or company specific workflows. They may also use stricter security policies, such as limited network access or custom base images. In these settings, the absolute success rates and costs may change. In addition, we run all experiments on a specific Linux distribution and hardware profile. Different operating systems, container runtimes or resource limits may create new failure modes.

The language and task coverage of AES-Bench is broad but still incomplete. For example, we do not include very large monorepos, data intensive pipelines, or projects with complex hardware dependencies. We also do not cover mobile or embedded platforms. The current results therefore best reflect medium to large server side and library style projects on common stacks. We expect the relative trends between EnvPilot and baselines to be stable on similar tasks, but we cannot claim full generality. Future work can reduce these threats by adding more domains, industrial case studies and larger repositories to the benchmark.

\textbf{Threats to internal validity.}
Internal validity concerns whether the observed effects are truly caused by EnvPilot.
We control many factors in our experiments. All methods run in the same container images, on the same hardware and under the same resource limits. We use shared tools, such as the environment checker and shell executor, across EnvPilot and baselines. We also reuse official prompts for existing agents whenever possible and follow their published settings. These steps reduce unfair advantages from implementation details.
Large language models are stochastic. To limit random variation, we fix model parameters and decoding settings for all systems. For each task, we run one setup attempt per method and record the full trajectory. Repeated trials would further reduce noise but are expensive at our scale. There is also a risk of subtle implementation bugs in our reimplementation of baselines. To reduce this risk, we follow their public code and perform manual checks on sampled trajectories.
We also take care when building the TDM. We construct TDM from external trajectories and use scripts to exclude projects that overlap with AES-Bench. This eliminates the risk that EnvPilot directly sees the target repositories in its memory. Still, some weaker forms of data leakage may remain. For example, the base LLMs may have seen related code fragments during pre-training. This limitation applies to all compared methods and is hard to avoid with current models. We therefore focus on relative improvements under the same model family and toolchain.

\textbf{Threats to construct validity.}
Construct validity concerns whether our metrics reflect the concepts that we want to study. Our main target is practical environment setup success. 
However, our success rate metric does not capture all aspects of usefulness. The current evaluation has a coarse granularity. It summarizes each task as success or failure and ignores many shades in between.
In practice, different projects and teams may adopt different definitions of a successful environment setup. Some settings require that all tests pass before work can continue. Other settings only require that key commands run and core functions execute without errors. Our benchmark does not distinguish these cases. It treats them as the same type of success or failure.
A more fine-grained benchmark could define several levels of success and report them separately. It could also separate functional readiness from full test conformance. Looking forward, we hope the community can design more systematic and widely accepted benchmarks for environment setup. We also plan to release trajectories and evaluation scripts to support independent replication and further refinement of metrics.

\section{Conclusions}
In this paper, \addedchw{we have presented the systematic design and empirical validation of EnvPilot, an experience-augmented agent tailored to software environment setup.} EnvPilot transforms execution trajectories into structured experience and uses the Context-aware Retrieval mechanism to guide environment setup. With this design, the agent learns from past tasks, avoids repeated errors, and adapts to diverse projects and languages.
We also introduced AES-Bench, a multilingual benchmark that contains 112 real-world instances across 9 programming languages. It provides a rigorous way to evaluate automated environment setup systems.
Experiments show that EnvPilot achieves an outstanding success rate of 75.00\%, while maintaining high efficiency. Ablation studies confirm that both structured experience and the Context-aware Retrieval mechanism are key to this performance. It shows that experience-driven methods can improve reliability and efficiency in automated environment setup. In future work, we plan to scale EnvPilot to larger repositories, integrate richer types of execution feedback, and explore collaboration with developers in real workflows.



\begin{acks}
This research is supported by National Natural Science Foundation of China (12201427) , Guangdong Basic and Applied Basic Research Foundation  (2021A1515110099) and Natural Science Foundation of Top Talent of SZTU (GDRC202133).
\end{acks}


\bibliographystyle{ACM-Reference-Format}
\bibliography{sample-base}

\newpage

\appendix

\section{Collections of Used Prompts}
\label{appendix:used_prompts}

\begin{lstlisting}[
  style=promptstyle,
  caption={Prompt of Instruction of Agent in automated environment setup Tasks},
]
You are an autonomous programmer, and you're working directly in the command line with a special interface.

In addition to typical bash commands, you can also use specific commands to help you navigate and edit files.
To call a command, you need to invoke it with a function call/tool call. 

RESPONSE FORMAT:
Your shell prompt is formatted as follows:
(Open file: <path>)
(Current directory: <cwd>)
bash-$

First, you should _always_ include a general thought about what you're going to do next.
Then, for every response, you must include exactly _ONE_ tool call/function call.

Remember, you should always include a _SINGLE_ tool call/function call and then wait for a response from the shell before continuing with more discussion and commands. Everything you include in the DISCUSSION section will be saved for future reference.
If you'd like to issue two commands at once, PLEASE DO NOT DO THAT! Please instead first submit just the first tool call, and then after receiving a response you'll be able to issue the second .
Note that the environment does NOT support interactive session commands (e.g. python, vim), so please do not invoke them.
instance_template: |-
We are currently configuring the environment for the repository. Your task is to configure the environment and installs this project (on an Ubuntu Linux machine) from source code and runs test cases.

{{Codebase_Profile}}

{{Prior_Knowledge}}

TASK TIPS:
1. It is prohibited to directly install this repository using dependency package management tools. For example, if current repository is Django, directly running pip install Django is strictly prohibited.
2. **PRIORITY ORDER for environment configuration discovery:**
  a) **First, check CI/CD configuration files** (.github/workflows/*.yml, .github/workflows/*.yaml, .gitlab-ci.yml, .circleci/config.yml, azure-pipelines.yml, Jenkinsfile, etc.) - these often contain the most reliable setup steps and test commands
  b) If no CI/CD files exist or they're insufficient, check dependency/environment files (requirements.txt, setup.py, pyproject.toml, package.json, Gemfile, Cargo.toml, etc.)
  c) Then examine README files (README.md, README.rst, etc.) for setup instructions
  d) Look for other configuration files (Makefile, tox.ini, environment.yml, etc.)
3. Always start by browsing the repository directory structure, with particular focus on the priority order above.
4. The choice of test framework should be determined by the repository's contents and CI/CD configurations.
5. It is strictly prohibited to modify the test cases in the code repository.
6. Do not create a new Dockerfile for environment isolation.
7. Commands can be run without sudo as the current session already has root privileges.
8. Based on the characteristics of the repository, it is possible to determine whether environment setup and test execution need to be performed within a virtual environment.
9. It may be necessary to install the repository from source before you can run code.
10. If you encounter package installation failures, try updating your local package index.
11. **Pay special attention to CI/CD matrix configurations** - they often test multiple Python/Node.js/etc. versions and can guide your environment setup.

GENERAL TIPS:
1. If you run a command and it doesn't work, try running a different command. A command that did not work once will not work the second time unless you modify it!
2. If you open a file and need to get to an area around a specific line that is not in the first 100 lines, say line 583, don't just use the scroll_down command multiple times. Instead, use the goto 583 command. It's much quicker.
3. Always make sure to look at the currently open file and the current working directory (which appears right after the currently open file). The currently open file might be in a different directory than the working directory! Note that some commands, such as 'create', open files, so they might change the current open file.
4. When using the edit command, always quote both search and replace arguments to avoid argument parsing failures.

INSTRUCTIONS:
Now, you're going to solve this issue on your own. Your terminal session has started and you're in the repository's root directory. You can use any bash commands or the special interface to help you. Edit all the files you need to and run any checks or tests that you want.
Remember, YOU SHOULD ALWAYS INCLUDE EXACTLY ONE TOOL CALL/FUNCTION CALL PER RESPONSE.
When you're satisfied with all of the changes you've made, you can submit your changes to the code base by simply running the submit command.
Note however that you cannot use any interactive session commands (e.g. python, vim) in this environment, but you can write scripts and run them. E.g. you can write a python script and then run it with the python command.
next_step_template: |-
{{observation}}
(Open file: {{open_file}})
(Current directory: {{working_dir}})
bash-$
next_step_no_output_template: |-
Your command ran successfully and did not produce any output.
(Open file: {{open_file}})
(Current directory: {{working_dir}})
bash-$
success_reflection_template: |-
You are an expert code assistant who just successfully repaired a software project in an automated programming session. Your task now is to reflect on this successful task and summarize key takeaways for future similar tasks.

CONTEXT:
Below is the full interaction history during the session, including commands issued, file edits, and observations returned. Use this history to extract useful insights.

----------------
{{session_history}}
----------------

PLEASE WRITE A REFLECTION INCLUDING:
1. **Problem Summary**:
  - What was the main problem or failure mode encountered?

2. **Critical Fix Steps**:
  - List 2-5 concrete steps that were essential to achieving success.
  - For each step, describe what the agent did and why it was effective.

3. **Heuristics or General Patterns**:
  - Are there any general rules or repeatable strategies that could help in future repairs?

4. **Environment or Tooling Insights** (if applicable):
  - Any key observations about the build system, environment configuration, or command-line tools that helped?

Here is the traj:
\end{lstlisting}

\begin{lstlisting}[
  style=promptstyle,
  caption={Prompt of Extract Codebase-related Info},
]
As an expert software engineer, your task is to analyze project files to extract the technical components required to set up its environment.
**IMPORTANT CONTEXT: You are operating inside a fresh, clean Linux container (e.g., Ubuntu22.04). Your goal is to determine the necessary commands and tools to install *directly within this existing container*.**

**CRITICAL CONSTRAINTS:**
- **DO NOT include Docker-related setup steps.** You are already inside a container, so ignore Dockerfile, docker build, docker run, or any containerization instructions.
- **Focus on what needs to be installed INSIDE the container**, not how to build or run containers.
- If you see a Dockerfile, extract the `RUN apt-get install ...` commands as system dependencies, but ignore `FROM`, `docker build`, or `docker run` instructions.

--- README.md ---
{readme}
--- END README.md ---

--- Workflow Guidelines (e.g., .github/workflows/ci.yml, Dockerfile) ---
{workflow_guide}
--- END WORKFLOW GUIDELINES ---

Based on the files, extract the key information into the following JSON schema.

**CONSTRAINTS:**
- **DO NOT include Docker-related setup steps.** Your analysis should focus on what needs to be done *inside* a container, not on how to build one. For example, if you see a Dockerfile, extract the `RUN apt-get install ...` commands as system dependencies, but ignore `FROM`, `docker build`, or `docker run` instructions.
- If a tool or dependency is mentioned, extract it. If not, use `null` or an empty list `[]`.

Fill the following JSON schema:
{{
    "language": {{
        "name": "string (e.g., Python, Node.js, C/C++)",
        "version_constraint": "string (e.g., 3.10, >=16.x)"
    }},
    "package_manager": {{
        "name": "string (e.g., pip, npm)",
        "install_command": "string (e.g., pip install -r requirements.txt)",
        "config_file": "string (e.g., pyproject.toml, package.json)"
    }},
    "build_tools": ["string (e.g., make, cmake, gradle)"],
    "test_framework": ["string (e.g., pytest, jest)"],
    "system_dependencies": {{
        "identified_in_ci": ["string (system packages from CI/workflow files, e.g., gcc, make)"],
        "likely_required": ["string (inferred but not explicitly listed system dependencies, e.g., libssl-dev)"]
    }},
}}
Return only the JSON object with the extracted information.
\end{lstlisting}

\begin{lstlisting}[
  style=promptstyle,
  caption={Prompt of Generate Hypothetical Query},
]
    As an experienced developer setting up a project, generate 3-5 hypothetical, specific questions you might ask when facing setup problems. Base your questions on the provided technical analysis. These questions should be perfect for retrieving solutions from a knowledge base.

    **IMPORTANT CONTEXT: You are setting up the environment INSIDE an existing Linux container. DO NOT include Docker-related questions or containerization setup.**

    --- Project Technology Stack Analysis ---
    {profile_str}
    --- END ANALYSIS ---
    
    The questions must be:
    1.  INTENT-ALIGNED: Address specific problems (e.g., "How to fix...").
    2.  CONTEXT-RICH: Include versions, OS, libraries from the analysis.
    3.  PROACTIVE: Predict common pitfalls like version conflicts or missing dependencies.
    4.  CONTAINER-AWARE: Focus on installing tools and dependencies inside the container, not on containerization itself.

    **AVOID questions about:**
    - Docker setup, container building, or containerization
    - "How to run this in Docker" or "How to build a container"
    - Container orchestration or deployment

    **FOCUS on questions about:**
    - Installing system packages with apt-get/yum
    - Setting up language runtimes and package managers
    - Configuring build tools and dependencies
    - Resolving version conflicts within the container
    
    Return only a JSON object with a "queries" key containing a list of question strings.
\end{lstlisting}

\begin{lstlisting}[
  style=promptstyle,
  caption={Prompt of Experience Extraction from Trajectories},
]
You are an expert in analyzing environment configuration problems, specializing in identifying and analyzing environment-related issues from agent trajectories.

## Objective
Based on the given agent trajectory, use **Chain of Thought** analysis process to accurately identify environment configuration problems and extract structured solutions.

## Analysis Steps
1. **Problem Identification**: Carefully read the trajectory to identify all environment configuration-related failures or errors
2. **Root Cause Analysis**: Deeply analyze the root cause of each problem
3. **Solution Extraction**: Extract actually effective solutions from the trajectory
4. **Structured Output**: Generate standardized triple format

## Key Requirements
- **Generality**: Descriptions should be **repository-agnostic**, focusing on general environment configuration problem types rather than being specific to a particular project
- **Completeness**: Ensure all environment configuration-related problems are identified and recorded
- **Accuracy**: Only record problems that actually occurred in the trajectory and solutions that were truly effective

## Output Format
Return a valid JSON object with the following structure:
{
    "triples": [
        {
            "problem": "Concise problem description (e.g., 'The initial make command failed when building the project')",
            "solution": "Detailed solution explanation including root cause analysis (e.g., 'The make tool was not installed on the Ubuntu system, which is required for building projects using Makefiles. The root cause was a missing build tool package in the environment configuration. The solution was to install make using apt-get install -y make')",
            "action": "The specific command used to solve the problem (e.g., 'apt-get install -y make')"
        }
    ]
}
**IMPORTANT**: Return ONLY valid JSON format, no additional text or explanations.
\end{lstlisting}

\begin{lstlisting}[
  style=promptstyle,
  caption={Prompt of LLM judge},
]
# Agent Environment Configuration Evaluation Task

## Role Definition
You are an expert Software Quality Assurance (QA) Judge. Your task is to evaluate whether an automated agent has successfully configured a software repository environment. Core Principle: You are judging the integrity of the environment, not the correctness of the code logic. The Test Lifecycle Completeness itself is the ultimate validater of the environment. Whether the test passes or fails reflects the correctness of the code logic.

## Key Definitions (Crucial)
Before evaluating, distinguish between these status types:
- FAILED (Assertion Failure): The test ran, but the result did not match the expectation (e.g., assert 1 == 2). This is ACCEPTABLE.
- ERROR (Runtime/Environment Error): The test could not run due to missing dependencies, syntax errors, or environment issues (e.g., ImportError, ModuleNotFound, ConnectionRefused caused by config). This is UNACCEPTABLE.
- CRASH (Process Abort): The test runner stopped unexpectedly (e.g., Segmentation Fault, process killed) before generating a final report. This is UNACCEPTABLE.

## Success Criteria (PASSED)
The environment configuration is PASSED if and only if the following logical flow is satisfied:

1. Compilation/Build Phase (If Applicable)
- Requirement: The core source code must compile.
- Special Logic for "Build + Test" Commands:
  - If the build command (e.g., mvn install, ./gradlew build) runs tests automatically and fails only because of Assertion Failures, but the compilation phase passed and a test report was generated, this counts as SUCCESS.
  - If the build fails due to compilation errors or missing dependencies, it is FAILED.

2. Test Execution Phase (Mandatory)
- Requirement: The test suite MUST attempt to run. If no tests are executed, it is FAILED.
- Stability: The testing process must complete a full lifecycle without crashing.
- Report Integrity: A valid final test report or summary line must be generated (e.g., "5 passed, 2 failed, 0 errors").

3. Final Verification (The "Golden Rule")
The configuration is successful if the logs show:
- Full Lifecycle: The test runner started and finished.
- Zero Errors: The test report shows 0 Errors.
- Allowed Failures: The test report can show > 0 Failures (Assertion faults).
- Report Generated: A summary confirms the test run is complete.

## Failure Scenarios (Critical)
The result is **FAILED** if ANY of the following occur:
1. **No Tests Run:** The agent set up the environment but never executed the test command.
2. **Process Crash:** The test command exited abruptly without producing a final summary/report.
3. **Environment/Runtime Errors:** The output contains `Error` status for tests (indicating environment issues rather than logic bugs) or tracebacks indicating missing modules/libraries.
4. **Compilation Error:** The actual code compilation failed (syntax errors, missing headers), distinct from test assertion failures.
5. **Incomplete Test Report:** The logs end mid-test without a final tally.
6. **Tests Not Found:** Since the target repository contains tests, if no tests are found and therefore the tests are not executed, it is also considered a failure.

## Evaluation Logic
1. **Did the agent run a test command?**
   - NO -> **FAILED**.
2. **Did the build/test process crash or hang?**
   - YES -> **FAILED**.
3. **Did the process output a complete test report/summary?**
   - NO -> **FAILED**.
4. **Analyze the Report/Output:**
   - Are there `Errors` (Runtime/Env issues)? -> **FAILED**.
   - Are there `Failures` (Assertions)? -> **PASSED** (Environment is fine, code is just buggy).
   - Is the exit code non-zero purely due to `Failures`? -> **PASSED**.

## Task Requirements
Based on the agent output trajectory I provide, judge whether the environment configuration meets the success criteria above.

Pay attention to the log of the last executed test or compilation command in the trajectory, and don't be distracted by the previous agent trial and error information (because the agent may have been configured incorrectly before finally being configured successfully).

## Output Format
1. First line: PASSED or FAILED (direct output, no other text)
2. Following lines: Detailed reasoning explaining your judgment based on the two criteria above
\end{lstlisting}

\section{\addedfinal{Baseline Adaptation Details: SWE-Agent for Environment Setup}}
\label{appendix:swe_agent_adaptation}
This appendix provides reproducible details for the \emph{``SWE-Agent (Adapted for Environment Setup)''} baseline used in Section~\ref{subsec:baselines}. The goal is to make the adaptation transparent and to ensure the comparison is fair.

\begin{table}[H]
  \centering
  \small
  \caption{Key differences between the original SWE-Agent and our adaptation for automated environment setup in AES-Bench.}
  \label{tab:swe_agent_adaptation}
  \setlength{\tabcolsep}{6pt}
  \begin{tabularx}{\linewidth}{lXX}
    \toprule
    \textbf{} & \textbf{SWE-Agent} & \textbf{Adapted SWE-Agent (Environment Setup)} \\
    \midrule
    Task Objective & Fix a repository-level bug by producing a code patch (Git diff) that makes tests pass. & Configure a reproducible environment and execute the test suite. \\
    Input & A GitHub issue/PR description (natural language) plus the target repository snapshot. & A repository URL plus a fixed commit SHA (version-locked), matching the AES-Bench instance definition (Section~\ref{sec:benchmark}). \\
    Output Artifact & A Git diff patch extracted and saved by the system for evaluation. &  A trajectory log, a test log and a setup script.\\
    Submit Behavior & Marks completion and returns the proposed patch. & Marks completion and indicates that the agent believes the environment setup is finished; evaluation is performed post-hoc from logs. \\
    Evaluation & Test pass rate (fail-to-pass) on SWE-bench style tasks. & LLM-based evaluation that checks test lifecycle completeness and log evidence. \\
    Prompt Adaptation & Oriented to code repair (diagnose bug and modify source code). & Oriented to environment setup (install dependencies, satisfy toolchain constraints, and run tests). \\
    \bottomrule
  \end{tabularx}
\end{table}

\textit{Tool access.} 
The adapted SWE-Agent uses the same tool interfaces and permissions as the original SWE-Agent implementation.

\textit{Stopping conditions.} 
We keep the stopping conditions identical to the original SWE-Agent, except that the success objective changes from producing a patch to completing environment setup and test execution.
Concretely, we cap the adapted SWE-Agent to the same maximum number of interaction turns as EnvPilot (50 turns, Section~\ref{subsec:implementation}). The run terminates early if the agent triggers \texttt{submit}.

\textit{Task-specific Heuristics.} 
We do not introduce any additional, benchmark-specific heuristics (e.g., hard-coded language-specific installers, curated dependency lists, or hand-written per-task scripts). The only adaptation is in the task specification and the prompt content to steer the agent toward environment setup.

\textit{Prompt adaptation.}
We provide the adapted prompt used by SWE-Agent in this baseline below. 

\begin{lstlisting}[
  style=promptstyle,
  caption={Prompt of adapted SWE-Agent for automated environment setup tasks.},
]
templates:
    system_template: |-
        SETTING: You are an autonomous programmer, and you're working directly in the command line with a special interface.

        In addition to typical bash commands, you can also use specific commands to help you navigate and edit files.
        To call a command, you need to invoke it with a function call/tool call. 

        RESPONSE FORMAT:
        Your shell prompt is formatted as follows:
        (Open file: <path>)
        (Current directory: <cwd>)
        bash-$

        First, you should _always_ include a general thought about what you're going to do next.
        Then, for every response, you must include exactly _ONE_ tool call/function call.

        Remember, you should always include a _SINGLE_ tool call/function call and then wait for a response from the shell before continuing with more discussion and commands. Everything you include in the DISCUSSION section will be saved for future reference.
        If you'd like to issue two commands at once, PLEASE DO NOT DO THAT! Please instead first submit just the first tool call, and then after receiving a response you'll be able to issue the second .
        Note that the environment does NOT support interactive session commands (e.g. python, vim), so please do not invoke them.
    instance_template: |-
        We are currently configuring the environment for the repository. Your task is to configure the environment and installs this project (on an Ubuntu Linux machine) from source code and runs test cases.

        {{repo_related_info}}

        {{cross_repo_knowledge}}

        TASK TIPS:
        1. It is prohibited to directly install this repository using dependency package management tools. For example, if current repository is Django, directly running pip install Django is strictly prohibited.
        2. **PRIORITY ORDER for environment configuration discovery:**
            a) **First, check CI/CD configuration files** (.github/workflows/*.yml, .github/workflows/*.yaml, .gitlab-ci.yml, .circleci/config.yml, azure-pipelines.yml, Jenkinsfile, etc.) - these often contain the most reliable setup steps and test commands
            b) If no CI/CD files exist or they're insufficient, check dependency/environment files (requirements.txt, setup.py, pyproject.toml, package.json, Gemfile, Cargo.toml, etc.)
            c) Then examine README files (README.md, README.rst, etc.) for setup instructions
            d) Look for other configuration files (Makefile, tox.ini, environment.yml, etc.)
        3. Always start by browsing the repository directory structure, with particular focus on the priority order above.
        4. The choice of test framework should be determined by the repository's contents and CI/CD configurations.
        5. It is strictly prohibited to modify the test cases in the code repository.
        6. Do not create a new Dockerfile for environment isolation.
        7. Commands can be run without sudo as the current session already has root privileges.
        8. Based on the characteristics of the repository, it is possible to determine whether environment setup and test execution need to be performed within a virtual environment.
        9. It may be necessary to install the repository from source before you can run code.
        10. If you encounter package installation failures, try updating your local package index.
        11. **Pay special attention to CI/CD matrix configurations** - they often test multiple Python/Node.js/etc. versions and can guide your environment setup.

        GENERAL TIPS:
        1. If you run a command and it doesn't work, try running a different command. A command that did not work once will not work the second time unless you modify it!
        2. If you open a file and need to get to an area around a specific line that is not in the first 100 lines, say line 583, don't just use the scroll_down command multiple times. Instead, use the goto 583 command. It's much quicker.
        3. Always make sure to look at the currently open file and the current working directory (which appears right after the currently open file). The currently open file might be in a different directory than the working directory! Note that some commands, such as 'create', open files, so they might change the current open file.
        4. When using the edit command, always quote both search and replace arguments to avoid argument parsing failures.

        INSTRUCTIONS:
        Now, you're going to solve this issue on your own. Your terminal session has started and you're in the repository's root directory. You can use any bash commands or the special interface to help you. Edit all the files you need to and run any checks or tests that you want.
        Remember, YOU SHOULD ALWAYS INCLUDE EXACTLY ONE TOOL CALL/FUNCTION CALL PER RESPONSE.
        When you're satisfied with all of the changes you've made, you can submit your changes to the code base by simply running the submit command.
        Note however that you cannot use any interactive session commands (e.g. python, vim) in this environment, but you can write scripts and run them. E.g. you can write a python script and then run it with the python command.
    next_step_template: |-
        {{observation}}
        (Open file: {{open_file}})
        (Current directory: {{working_dir}})
        bash-$
    next_step_no_output_template: |-
        Your command ran successfully and did not produce any output.
        (Open file: {{open_file}})
        (Current directory: {{working_dir}})
        bash-$
    success_reflection_template: |-
        You are an expert code assistant who just successfully repaired a software project in an automated programming session. Your task now is to reflect on this successful task and summarize key takeaways for future similar tasks.

        CONTEXT:
        Below is the full interaction history during the session, including commands issued, file edits, and observations returned. Use this history to extract useful insights.

        ----------------
        {{session_history}}
        ----------------

        PLEASE WRITE A REFLECTION INCLUDING:
        1. **Problem Summary**:
            - What was the main problem or failure mode encountered?

        2. **Critical Fix Steps**:
            - List 2-5 concrete steps that were essential to achieving success.
            - For each step, describe what the agent did and why it was effective.

        3. **Heuristics or General Patterns**:
            - Are there any general rules or repeatable strategies that could help in future repairs?

        4. **Environment or Tooling Insights** (if applicable):
            - Any key observations about the build system, environment configuration, or command-line tools that helped?

        Here is the traj:
\end{lstlisting}

\section{\addedfinal{Complete List of AES-Bench Instances}}
\label{appendix:dataset_instances}

To ensure full reproducibility, we provide the complete list of all 112 instances in AES-Bench. Each instance is identified by its repository and commit hash, grouped by programming language.

\subsection{C Language Instances (12 instances)}
\begin{longtable}{@{}>{\raggedright\arraybackslash}p{\dimexpr .30\linewidth-2\tabcolsep\relax}>{\raggedright\arraybackslash}p{\dimexpr .10\linewidth-2\tabcolsep\relax}>{\raggedright\arraybackslash}p{.60\linewidth}@{}}
\toprule
\textbf{Repository} & \textbf{Pull \#} & \textbf{Base Commit} \\
\midrule
\endhead
fluent/fluent-bit & 3663 & b0f0b290375ecac2e3b1979ffb4b42331d58367a \\
fluent/fluent-bit & 10007 & 9f2502b99db98e26efe9ac27bb0cf43a453f108b \\
libsdl-org/SDL & 11946 & f731741eadbe06d618d6a9bbdb41c2ee98b8178d \\
mruby/mruby & 3649 & ef305305b43528d94b162078c421a0590ba1c438 \\
mruby/mruby & 6442 & 4a99f28ec37152babd82328b1441b5cca6a311d5 \\
OpenMathLib/OpenBLAS & 4729 & 1ba1b9c357fb7d5916a56a6c388b4aea47aad395 \\
php/php-src & 17835 & e8dda54dd58d0792185916da4d65a8c544125486 \\
redis/redis & 13711 & dc57ee03b1c5b8f646718e362f3a809a7511ad36 \\
valkey-io/valkey & 1694 & 1470ee8a639fea32ae3cfcb8b12e398f69eb1f83 \\
mpv-player/mpv & 16604 & 03cafa10a72f9b8aa458621c3afb9695089014d6 \\
taosdata/TDengine & 32431 & ee078f0a17d159f500d6e6296aa549db8d99c089 \\
obsproject/obs-studio & 12432 & e7570391067f82e3511ed1b31a76d758064cc790 \\
\bottomrule
\end{longtable}

\subsection{C++ Language Instances (13 instances)}
\begin{longtable}{@{}>{\raggedright\arraybackslash}p{\dimexpr .30\linewidth-2\tabcolsep\relax}>{\raggedright\arraybackslash}p{\dimexpr .10\linewidth-2\tabcolsep\relax}>{\raggedright\arraybackslash}p{.60\linewidth}@{}}
\toprule
\textbf{Repository} & \textbf{Pull \#} & \textbf{Base Commit} \\
\midrule
\endhead
CGAL/cgal & 8736 & 8675d06ea197478d7687b5d52307e1c4cf9c2ada \\
halide/Halide & 7506 & e7f78600e10956b44e8f214c686f310211b0d836 \\
halide/Halide & 6533 & cf8c8f22eb507aedeba5a44f8ee20bc63757dc57 \\
halide/Halide & 5545 & 392b53eb22b8edfb69de5dc27d16dca5fc7512a7 \\
halide/Halide & 8490 & 166cd92b2100f3bbde989cbe26def7063c9c8e0a \\
root-project/root & 17731 & bfeb0a10add47ff5ba09dfc9ea723aa8d0665106 \\
godotengine/godot & 109181 & cce10e0b22a6323fbafa5082d2105091b6d193eb \\
ggml-org/llama.cpp & 14934 & 4cb208c93c1c938591a5b40354e2a6f9b94489bc \\
dragonflydb/dragonfly & 5598 & a895afc971597b9168f0e814d4ea4b7e54dbd8aa \\
taichi-dev/taichi & 8752 & ba0e81dce559fb63a5958bf82feb1d00c55c02fe \\
rethinkdb/rethinkdb & 7166 & a6c7cbc345dac1eac86fe343a380cb14cd2a2980 \\
google/flatbuffers & 8649 & 518bf42df82da02b77696027215bda91edbd4102 \\
google/libphonenumber & 3877 & 5e35f6ea56c60ee1f7dbffe67a1eaaff8dfacda3 \\
\bottomrule
\end{longtable}

\subsection{Go Language Instances (13 instances)}
\begin{longtable}{@{}>{\raggedright\arraybackslash}p{\dimexpr .30\linewidth-2\tabcolsep\relax}>{\raggedright\arraybackslash}p{\dimexpr .10\linewidth-2\tabcolsep\relax}>{\raggedright\arraybackslash}p{.60\linewidth}@{}}
\toprule
\textbf{Repository} & \textbf{Pull \#} & \textbf{Base Commit} \\
\midrule
\endhead
beego/beego & 5725 & ff9fedc989ba7eb5413d3ab28225f14948d1a33b \\
caddyserver/caddy & 6669 & 350ad38f63f7a49ceb3821c58d689b85a27ec4e5 \\
fatedier/frp & 3664 & df12cc2b9d24eb0cbf0b64078c97bf32d8c2da9c \\
gin-gonic/gin & 4048 & 28e57f58b184b2305ace192e02496bb89f6fd8cb \\
go-gorm/gorm & 5974 & 2bc913787b6d194aa4f72c8e4ddc64d62602ef21 \\
gohugoio/hugo & 13495 & 61c39ae63b62667d965c2ff96d085f4eda59bcb2 \\
istio/istio & 55229 & fbe084f46ed796559f0d0afdbb5a916988b07454 \\
jesseduffield/lazygit & 4369 & 19ac926116863740e63cf7dabcb46845c8d41b91 \\
junegunn/fzf & 4231 & 2b584586ed1caf15429625da981575ee35d407b8 \\
labstack/echo & 2717 & 3b017855b4d331002e2b8b28e903679b875ae3e9 \\
prometheus/prometheus & 16195 & b0227d1f16ea5da448f7a610ed9a7e22e6f35782 \\
syncthing/syncthing & 9914 & ab20c16982b1fa4cc2d84c3f88db054d9b21793b \\
concourse/concourse & 9103 & 55ff4b2410cb67746c14438a0af658151deb20b0 \\
\bottomrule
\end{longtable}

\subsection{Java Language Instances (13 instances)}
\begin{longtable}{@{}>{\raggedright\arraybackslash}p{\dimexpr .30\linewidth-2\tabcolsep\relax}>{\raggedright\arraybackslash}p{\dimexpr .10\linewidth-2\tabcolsep\relax}>{\raggedright\arraybackslash}p{.60\linewidth}@{}}
\toprule
\textbf{Repository} & \textbf{Pull \#} & \textbf{Base Commit} \\
\midrule
\endhead
checkstyle/checkstyle & 15448 & d68addb533f2f8934b5a12b4b4a57a1215672b2d \\
junit-team/junit5 & 4366 & 2e27ea70bd9ad30049a922bce882719e5e726f81 \\
junit-team/junit5 & 3394 & ba34eed48691f8c076674e25266ee4f9254356c6 \\
apache/shenyu & 5869 & 5b8e7d07f68e36d4457f354d57b4cbf5411ef817 \\
Graylog2/graylog2-server & 22093 & ccd21485213ba3d9a107c7f5988a2e1dc5dc2b37 \\
Graylog2/graylog2-server & 14865 & 851d870354e9b9bcef5c7f2b6ecfd6752221ae92 \\
micronaut-projects/micronaut-core & 11440 & 40478d20ba4430dbedf5e24a26d580e1b1ceeae9 \\
OpenRefine/OpenRefine & 7174 & ced412aca3422d7d46f26913b33217d0af20e71c \\
provectus/kafka-ui & 3505 & 4d03802a5dc77f6b327485484eb63106a87152b5 \\
google/gson & 2885 & f7de5c2c22dae98f2f98eefcb70c920ff71e437b \\
openjdk/jdk & 26541 & fe09e93b8fc3081c944f3824fdaa55cc17e377a8 \\
eclipse-vertx/vert.x & 5650 & 12a0fd3a4dff0613c88bafafed632b8b19515d58 \\
projectlombok/lombok & 3886 & b223db18182c312b2ffac379d954ea9b6fd28b7e \\
\bottomrule
\end{longtable}

\subsection{JavaScript Language Instances (12 instances)}
\begin{longtable}{@{}>{\raggedright\arraybackslash}p{\dimexpr .30\linewidth-2\tabcolsep\relax}>{\raggedright\arraybackslash}p{\dimexpr .10\linewidth-2\tabcolsep\relax}>{\raggedright\arraybackslash}p{.60\linewidth}@{}}
\toprule
\textbf{Repository} & \textbf{Pull \#} & \textbf{Base Commit} \\
\midrule
\endhead
Automattic/mongoose & 14692 & 87c57e60bfe53cfa1a6cbd18727a075e9e1e8462 \\
Automattic/mongoose & 15302 & ae6f4a27235fc2368da09392d3e4d3651a56624b \\
caolan/async & 1224 & ca22b96ec512699dee73d252ec4f557d4cf9c31d \\
caolan/async & 1790 & 6ae4aaafa3541f0a5179e97d8263af9bd53d97eb \\
google/zx & 811 & 23fe548b8ecd69293246b2827ca8c7b6cb154315 \\
google/zx & 1113 & f4b0328bcb3e0718b048aaf29554a26536a4c077 \\
tj/commander.js & 987 & 80d80af65ffb45f56478d7f01f1f6316c50817d5 \\
tj/commander.js & 1926 & d0385706a17f183d86bf2a4b11b5d110aa3791a5 \\
facebook/react & 30951 & d6cb4e771341ff82489c00f4907990cb8a75696b \\
KaTeX/KaTeX & 3735 & be079843132408da1c3bf04fd6ebdd73da899cf0 \\
nasa/openmct & 7986 & 1fde0d9e38ba039ab2e531c5f508d5fe6b6dd169 \\
swagger-api/swagger-ui & 10390 & ac4b549bfdc36827aa6318f2c322243e4259e902 \\
\bottomrule
\end{longtable}

\subsection{PHP Language Instances (12 instances)}
\begin{longtable}{@{}>{\raggedright\arraybackslash}p{\dimexpr .30\linewidth-2\tabcolsep\relax}>{\raggedright\arraybackslash}p{\dimexpr .10\linewidth-2\tabcolsep\relax}>{\raggedright\arraybackslash}p{.60\linewidth}@{}}
\toprule
\textbf{Repository} & \textbf{Pull \#} & \textbf{Base Commit} \\
\midrule
\endhead
briannesbitt/Carbon & 3170 & e4e6956a47973a1787d1c50bb7ac4ce7cc738cfd \\
composer/composer & 12367 & 5e890f0dc8334219844911bff79a1e26036ad2bb \\
doctrine/dbal & 6890 & 790a5f0d4646d3e21817949876484d36a78db9e0 \\
laravel/framework & 55507 & 5dea84f6fdcb2c5826f8a3ac77f7818a2c3c6f29 \\
slimphp/Slim & 3319 & 038fd5713d5a41636fdff0e8dcceedecdd17fc17 \\
filamentphp/filament & 17110 & 859a37ce686bbea9e5cb7ba9bc21425c0b0cd84e \\
monicahq/monica & 7781 & d74034fd492d44225fe676dc2e8d5933b34c72f4 \\
PHPMailer/PHPMailer & 3170 & a2fa1021046c4a7bca84e300d5268026949a06bf \\
bcit-ci/CodeIgniter & 6261 & ae7b30f9cbaf462a3cc039de6d1e7738c6bea2d4 \\
nikic/PHP-Parser & 1088 & 7fc3bcf97001c4a7aae5c355f0b03e7af54ca728 \\
getgrav/grav & 3886 & 2620e836d4ac0ad495173efdfe98d0a2db94f92c \\
barryvdh/laravel-ide-helper & 1688 & cff244b9b5efbf155f5eceb1b2b1002425de751f \\
\bottomrule
\end{longtable}

\subsection{Python Language Instances (12 instances)}
\begin{longtable}{@{}>{\raggedright\arraybackslash}p{\dimexpr .30\linewidth-2\tabcolsep\relax}>{\raggedright\arraybackslash}p{\dimexpr .10\linewidth-2\tabcolsep\relax}>{\raggedright\arraybackslash}p{.60\linewidth}@{}}
\toprule
\textbf{Repository} & \textbf{Pull \#} & \textbf{Base Commit} \\
\midrule
\endhead
astropy/astropy & 13033 & 298ccb478e6bf092953bca67a3d29dc6c35f6752 \\
django/django & 10097 & b9cf764be62e77b4777b3a75ec256f6209a57671 \\
matplotlib/matplotlib & 13989 & a3e2897bfaf9eaac1d6649da535c4e721c89fa69 \\
mwaskom/seaborn & 3187 & 22cdfb0c93f8ec78492d87edb810f10cb7f57a31 \\
pallets/flask & 5014 & 7ee9ceb71e868944a46e1ff00b506772a53a4f1d \\
pydata/xarray & 4629 & a41edc7bf5302f2ea327943c0c48c532b12009bc \\
pytest-dev/pytest & 5262 & 58e6a09db49f34886ff13f3b7520dd0bcd7063cd \\
psf/requests & 1142 & 22623bd8c265b78b161542663ee980738441c307 \\
pylint-dev/pylint & 4970 & 40cc2ffd7887959157aaf469e09585ec2be7f528 \\
scikit-learn/scikit-learn & 10297 & b90661d6a46aa3619d3eec94d5281f5888add501 \\
sphinx-doc/sphinx & 10673 & f35d2a6cc726f97d0e859ca7a0e1729f7da8a6c8 \\
sympy/sympy & 12096 & d7c3045115693e887bcd03599b7ca4650ac5f2cb \\
\bottomrule
\end{longtable}

\subsection{Rust Language Instances (12 instances)}
\begin{longtable}{@{}>{\raggedright\arraybackslash}p{\dimexpr .30\linewidth-2\tabcolsep\relax}>{\raggedright\arraybackslash}p{\dimexpr .10\linewidth-2\tabcolsep\relax}>{\raggedright\arraybackslash}p{.60\linewidth}@{}}
\toprule
\textbf{Repository} & \textbf{Pull \#} & \textbf{Base Commit} \\
\midrule
\endhead
alacritty/alacritty & 8192 & c899208e15c8987480912794d66358ea2581fc39 \\
fish-shell/fish-shell & 10472 & f551eeadfeb90e68e5ed250b89ab8bcabf6caba0 \\
helix-editor/helix & 6675 & 76825c7b661f4526856627bf1e34f024088f3fda \\
rust-lang/mdBook & 2486 & 90960126e8b2da1dc1a7e460c2bdeb8073f1b81e \\
typst/typst & 6652 & e9f1b5825a9d37ca0c173a7b2830ba36a27ca9e0 \\
casey/just & 2836 & a3da3cf959a4cf9eba85777c4e5c0a95ad0e0d0b \\
rwf2/Rocket & 2838 & de6632ea56cce497d23a8f2223b8f8721202ff39 \\
astral-sh/uv & 14972 & d867f3e595189441a80f02b013ab09e571e12f35 \\
FuelLabs/sway & 7300 & 8b29cc39c52cf7a3fd7857290f9741334923d362 \\
vercel/turborepo & 10710 & f60418867da7304f83d2ce8168acfc917e687417 \\
ajeetdsouza/zoxide & 1080 & 6324b4e347a45ba7e07ec2584d11e133fad3bd84 \\
iced-rs/iced & 3019 & d4f8e1902322b239be9580b3afb936c99ec49b79 \\
\bottomrule
\end{longtable}

\subsection{TypeScript Language Instances (13 instances)}
\begin{longtable}{@{}>{\raggedright\arraybackslash}p{\dimexpr .30\linewidth-2\tabcolsep\relax}>{\raggedright\arraybackslash}p{\dimexpr .10\linewidth-2\tabcolsep\relax}>{\raggedright\arraybackslash}p{.60\linewidth}@{}}
\toprule
\textbf{Repository} & \textbf{Pull \#} & \textbf{Base Commit} \\
\midrule
\endhead
colinhacks/zod & 3887 & 207205cab4b4f33f9274a9cce5040d6ced692300 \\
trpc/trpc & 5532 & cdf215127a172e497468dec04acb8ea12e840e84 \\
trpc/trpc & 5850 & fbb4a2fb4b71d2f5ab7ee97617d951f3a92517c8 \\
nuxt/nuxt & 24511 & e3b8b84a247b8873d2443bcfdd23cf7615078f4b \\
nuxt/nuxt & 31081 & 88b119a27f89090a2d334298c46e456be9e54b2b \\
reduxjs/redux & 4519 & 461b0932cf66112105df862d7c018e7488f9f486 \\
Automattic/wp-calypso & 101046 & 4211bad2f5325ece5803e79785fa0eb8c8b87c82 \\
facebook/lexical & 6834 & d9014f328a92a3486fef2b345555c9772f1ec53c \\
MetaMask/metamask-extension & 30527 & 16dfc37f13e177cc79882353b15c3ff0a2fef08d \\
palantir/blueprint & 6882 & 1a5c82bcdfec804b9b65d44e4ac75b3af9aafe1e \\
payloadcms/payload & 12489 & 07e9444c09a418615e1366150f399048617ba5c1 \\
react-hook-form/react-hook-form & 12793 & 8e6dc9e65aab0c7e6c59ce6222f36f64b8645bcc \\
rjsf-team/react-jsonschema-form & 4637 & c05dbc19400b332c1a610bab68a178741812c227 \\
\bottomrule
\end{longtable}

\section{\addedfinal{Cross Language Analysis}}
\label{sec:cross_language_analysis}

\addedfinal{This section provides a language-wise breakdown of the aggregated results reported in Tab.~\ref{tab:main_results}. Our goal is to examine whether the advantage of EnvPilot is consistent across heterogeneous build systems and programming ecosystems, and to identify language-specific failure patterns.}

\noindent \addedfinal{\textbf{AES-Bench (112 tasks).} With DeepSeek-V3-0324 as the LLM, EnvPilot achieves the best overall ECSR (\textbf{75.00\%, 84/112}), while also reducing API calls to \textbf{28} with \textbf{220K} tokens and \textbf{\$0.063} cost per task (Tab.~\ref{tab:main_results}). The gains are broad and consistent across most languages: EnvPilot reaches \textbf{84.62\%} in Java (11/13), 84.62\% in Go (11/13), 80.00\% in JS/TS (20/25), 83.33\% in Rust (10/12), and 91.67\% in PHP (11/12). These results indicate that the same TDM schema and retrieval pipeline can generalize across languages with different toolchains (e.g., Maven/Gradle, Go modules and npm.).}

\addedfinal{A key takeaway is EnvPilot’s robustness on long-tail languages and heterogeneous dependency ecosystems. On PHP, EnvPilot achieves 91.67\% under DeepSeek-V3 and remains at 50.00\% under the smaller GPT-4o-mini, while RepoLaunch fails to complete any build (0.00\%) under GPT-4o-mini. This supports our hypothesis that TDM serves as a dynamic external knowledge base: it compensates for missing domain-specific setup knowledge in smaller models by retrieving expert knowledge.}

\addedfinal{We also observe language-specific exceptions that reveal future improvement opportunities. For C/C++, EnvPilot reaches 56.00\% (14/25), which is competitive with RepoLaunch (56.00\%) but below SWE-Agent (64.00\%). This suggests that C/C++ repositories exhibit highly heterogeneous toolchains and subtle environment differences, increasing the risk of misapplying retrieved experiences. In contrast, for JS/TS, EnvPilot is strong (80.00\%) but does not exceed RepoLaunch (80.00\%), indicating that for some ecosystems (e.g., npm-based), strong baseline heuristics may already capture many common setup patterns.}

\addedfinal{\textbf{AES-Bench under GPT-4o-mini.} When switching to GPT-4o-mini, the overall success rate of all methods decreases, yet EnvPilot still achieves the best total score (46.43\%, 52/112) with the lowest cost among learning-based baselines (0.034). The language-wise breakdown shows that TDM mitigates the small-model limitation in Go (46.15\%), PHP (50.00\%), and Rust (\textbf{66.67\%}), where setup errors are often driven by toolchain-specific dependencies rather than general reasoning alone. Meanwhile, for Java (53.85\%) and C/C++ (20.00\%), RepoLaunch remains competitive or better, implying that model capacity and stronger built-in heuristics can dominate when retrieved experience is less transferable.}

\addedfinal{\textbf{ExecutionAgent Bench (50 tasks).}Tab.~\ref{tab:main_results_ea} shows a similar cross-language trend on a different benchmark. Under DeepSeek-V3, EnvPilot achieves the best total success rate (72.00\%, 36/50) with 26 API calls, 198K tokens, and \$0.057 cost. The improvements are visible in multiple languages: C/C++ rises to 72.22\% (13/18) compared to 55.56\% for RepoLaunch and SWE-Agent; JS/TS reaches 72.73\% (8/11) compared to 45.45\% for RepoLaunch; and Python reaches 63.64\% (7/11), outperforming RepoLaunch (54.55\%) and SWE-Agent (36.36\%). These results suggest that TDM not only transfers across languages, but can also stabilize execution on a benchmark with different task composition and tooling constraints.}

\addedfinal{\textbf{ExecutionAgent Bench under GPT-4o-mini.} With GPT-4o-mini, EnvPilot remains the best overall method (46.00\%, 23/50). The advantage mainly comes from strong performance in toolchain-sensitive languages such as C/C++ (55.56\%, 10/18) and Python (36.36\%, 4/11), where retrieved experience can directly reduce trial-and-error. We note that for JS/TS, EnvPilot (36.36\%) lags behind RepoLaunch and SWE-Agent (both 45.45\%), suggesting that for certain ecosystems, retrieval quality (and experience transferability) becomes the bottleneck under smaller models.}

\addedfinal{\textbf{Summary.} Across two benchmarks and two backbone models, EnvPilot demonstrates consistent cross-language generalization: it achieves the best overall results and remains robust in long-tail languages (e.g., PHP and Rust). The few exceptions (notably C/C++ and some JS/TS settings) highlight that toolchain heterogeneity and negative transfer are the main challenges, motivating stronger context inference and experience filtering for highly diverse ecosystems.}

\begin{table}[htbp]
  \centering
  \small  
  \setlength{\tabcolsep}{4pt}  
  \caption{Main results of EnvPilot and baselines on \textbf{AES-Bench}. We highlight the \textbf{best} and \underline{second best} results.}
  \label{tab:main_results}
  \begin{adjustbox}{max width=\linewidth}
  \begin{tabular}{lllccc|lccc}
    \toprule
    \multirow{2}{*}{\textbf{Method}} & 
    \multirow{2}{*}{\textbf{Dataset}} & 
    \multicolumn{4}{c}{\textbf{DeepSeek-V3-0324}} & 
    \multicolumn{4}{c}{\textbf{GPT-4o-mini}} \\
    \cmidrule(lr){3-6}\cmidrule(lr){7-10}
    & & \textbf{Success (\%)} $\uparrow$ & \textbf{API} $\downarrow$ & \textbf{Token (K)} $\downarrow$ & \textbf{Cost (\$)} $\downarrow$
      & \textbf{Success (\%)} $\uparrow$ & \textbf{API} $\downarrow$ & \textbf{Token (K)} $\downarrow$ & \textbf{Cost (\$)} $\downarrow$ \\
    \midrule
    \multirow{8}{*}{ExecutionAgent} 
        & Total   & 43.75 \textcolor{gray}{(49/112)} & \textbf{28} & 420 & 0.130
                  & 23.21 \textcolor{gray}{(26/112)} & 44 & 643 & 0.105 \\
        & Python & 58.33 \textcolor{gray}{(7/12)}   & \underline{24} & 288 & 0.092
                  & \underline{41.67} \textcolor{gray}{(5/12)}   & 47 & 585 & 0.101 \\
        & C/C++  & 32.00 \textcolor{gray}{(8/25)}   & \underline{31} & 571 & 0.178
                  & 12.00 \textcolor{gray}{(3/25)}   & 42 & 792 & 0.128 \\
        & Java   & 30.77 \textcolor{gray}{(4/13)}   & \underline{27} & 420 & 0.135
                  & 15.38 \textcolor{gray}{(2/13)}   & 44 & 650 & 0.105 \\
        & JS/TS  & 44.00 \textcolor{gray}{(11/25)}  & \textbf{23} & 311 & 0.096
                  & 12.00 \textcolor{gray}{(3/25)}   & 46 & 618 & 0.101 \\
        & Go     & 53.85 \textcolor{gray}{(7/13)}   & 33 & 454 & 0.139
                  & \underline{15.38} \textcolor{gray}{(2/13)}   & 46 & 596 & 0.096 \\
        & PHP    & 41.67 \textcolor{gray}{(5/12)}   & \textbf{28} & 578 & 0.172
                  & \textbf{50.00} \textcolor{gray}{(6/12)}   & 41 & 552 & 0.089 \\
        & Rust   & 58.33 \textcolor{gray}{(7/12)}   & \textbf{20} & 268 & 0.082
                  & \underline{41.67} \textcolor{gray}{(5/12)}   & 42 & 584 & 0.094 \\
    \midrule
    \multirow{2}{*}{Repo2Run}
        & Total  & \underline{66.66} \textcolor{gray}{(8/12)} & \textbf{20} & 1000 & 0.281
                  & 25.00 \textcolor{gray}{(3/12)} & 45 & 1273 & 0.193 \\
        & Python & \textbf{66.66} \textcolor{gray}{(8/12)} & \textbf{20} & 1000 & 0.281
                  & 25.00 \textcolor{gray}{(3/12)} & 45 & 1273 & 0.193 \\
    \midrule
    \multirow{8}{*}{RepoLaunch} 
        & Total   & 55.36 \textcolor{gray}{(62/112)} & 34 & 250 & 0.071
                  & \underline{40.18} \textcolor{gray}{(45/112)} & 44 & 387 & 0.060 \\
        & Python & \textbf{66.67} \textcolor{gray}{(8/12)}   & 27 & \textbf{181} & \textbf{0.052}
                  & 33.33 \textcolor{gray}{(4/12)}   & 39 & 339 & 0.052 \\
        & C/C++  & 56.00 \textcolor{gray}{(14/25)}  & 40 & 358 & 0.102
                  & \textbf{32.00} \textcolor{gray}{(8/25)}   & 42 & 397 & 0.061 \\
        & Java   & 38.46 \textcolor{gray}{(5/13)}   & 28 & \underline{205} & \underline{0.059}
                  & \textbf{61.54} \textcolor{gray}{(8/13)}   & 40 & 282 & 0.044 \\
        & JS/TS  & \textbf{80.00} \textcolor{gray}{(20/25)} & 38 & 231 & 0.066
                  & \textbf{60.00} \textcolor{gray}{(15/25)} & 57 & 560 & 0.086 \\
        & Go     & \underline{76.92} \textcolor{gray}{(10/13)}   & 30 & 183 & 0.052
                  & \underline{15.38} \textcolor{gray}{(2/13)}   & 49 & 393 & 0.060 \\
        & PHP    & 0.00 \textcolor{gray}{(0/12)}   & \underline{29} & \underline{170} & \underline{0.049}
                  & 0.00 \textcolor{gray}{(0/12)}   & 30 & 173 & 0.026 \\
        & Rust   & 41.67 \textcolor{gray}{(5/12)}   & 39 & 332 & 0.095
                  & \textbf{66.67} \textcolor{gray}{(8/12)}   & 43 & 377 & 0.058 \\
    \midrule
    \multirow{8}{*}{SWE-Agent}
        & Total   & \underline{62.50} \textcolor{gray}{(70/112)} & \underline{30} & \textbf{200} & \textbf{0.058}
                  & 24.11 \textcolor{gray}{(27/112)} & \underline{29} & \textbf{192} & \textbf{0.030} \\
        & Python  & 50.00 \textcolor{gray}{(6/12)}   & 38 & 297 & 0.086
                  & 16.67 \textcolor{gray}{(2/12)}   & 31 & 173 & 0.027 \\
        & C/C++   & \textbf{64.00} \textcolor{gray}{(16/25)}  & \underline{31} & \textbf{206} & \textbf{0.059}
                  & \underline{24.00} \textcolor{gray}{(6/25)}   & 30 & 213 & 0.034 \\
        & Java    & \underline{61.54} \textcolor{gray}{(8/13)}   & 34 & 221 & 0.063
                  & 38.46 \textcolor{gray}{(5/13)}   & 26 & 170 & 0.027 \\
        & JS/TS   & \underline{76.00} \textcolor{gray}{(19/25)}  & 27 & \underline{170} & \underline{0.049}
                  & 24.00 \textcolor{gray}{(6/25)}   & 27 & 171 & 0.027 \\
        & Go      & 46.15 \textcolor{gray}{(6/13)}   & \underline{29} & \textbf{170} & \textbf{0.049}
                  & 0.00 \textcolor{gray}{(0/13)}    & 28 & 177 & 0.028 \\
        & PHP     & \underline{75.00} \textcolor{gray}{(9/12)}   & 34 & 209 & 0.060
                  & 41.67 \textcolor{gray}{(5/12)}   & 32 & 264 & 0.042 \\
        & Rust    & \underline{66.67} \textcolor{gray}{(8/12)}   & \underline{26} & \textbf{154} & \textbf{0.045}
                  & 25.00 \textcolor{gray}{(3/12)}   & 27 & 176 & 0.028 \\
    \midrule
    \multirow{8}{*}{EnvPilot (Ours)} 
        & Total   & \textbf{75.00} \textcolor{gray}{(84/112)} & \underline{28} & \underline{220} & \underline{0.063} 
                  & \textbf{46.43} \textcolor{gray}{(52/112)} & \textbf{26} & \underline{214} & \underline{0.034} \\
        & Python  & \textbf{66.67} \textcolor{gray}{(8/12)}   & 32 & 251 & 0.072 
                  & \textbf{41.67} \textcolor{gray}{(5/12)}  & \textbf{29} & 254 & 0.040 \\
        & C/C++   & \underline{56.00} \textcolor{gray}{(14/25)}  & \textbf{29} & \underline{238} & \underline{0.068} 
                  & 20.00 \textcolor{gray}{(5/25)}   & \textbf{29} & 251 & 0.039 \\
        & Java    & \textbf{84.62} \textcolor{gray}{(11/13)}  & \textbf{25} & \textbf{207} & \textbf{0.060} 
                  & \textbf{53.85} \textcolor{gray}{(7/13)}  & \underline{21} & 178 & 0.028 \\
        & JS/TS   & \textbf{80.00} \textcolor{gray}{(20/25)}  & \underline{24} & \textbf{168} & \textbf{0.048} 
                  & \textbf{60.00} \textcolor{gray}{(15/25)} & \textbf{25} & 199 & 0.031 \\
        & Go      & \textbf{84.62} \textcolor{gray}{(11/13)}  & \textbf{26} & \underline{186} & \underline{0.053} 
                  & \textbf{46.15} \textcolor{gray}{(6/13)}  & \textbf{26} & 176 & 0.027 \\
        & PHP     & \textbf{91.67} \textcolor{gray}{(11/12)}  & 34 & 275 & 0.079 
                  & \textbf{50.00} \textcolor{gray}{(6/12)}  & 32 & 270 & 0.042 \\
        & Rust    & \textbf{83.33} \textcolor{gray}{(10/12)}  & 29 & \underline{228} & \underline{0.066} 
                  & \textbf{66.67} \textcolor{gray}{(8/12)}  & \textbf{21} & \textbf{157} & \textbf{0.025} \\
  \bottomrule
  \end{tabular}
  \end{adjustbox}
\end{table}

\begin{table}[htbp]
  \centering
  \small
  \setlength{\tabcolsep}{4pt}
  \caption{Main results of EnvPilot and baselines on \textbf{ExecutionAgent Bench}. We highlight the \textbf{best} and \underline{second best} results.}
  \label{tab:main_results_ea}
  \begin{adjustbox}{max width=\linewidth}
  \begin{tabular}{lllccc|lccc}
    \toprule
    \multirow{2}{*}{\textbf{Method}} & 
    \multirow{2}{*}{\textbf{Dataset}} & 
    \multicolumn{4}{c}{\textbf{DeepSeek-V3-0324}} & 
    \multicolumn{4}{c}{\textbf{GPT-4o-mini}} \\
    \cmidrule(lr){3-6}\cmidrule(lr){7-10}
    & & \textbf{Success (\%)} $\uparrow$ & \textbf{API} $\downarrow$ & \textbf{Token (K)} $\downarrow$ & \textbf{Cost (\$)} $\downarrow$
      & \textbf{Success (\%)} $\uparrow$ & \textbf{API} $\downarrow$ & \textbf{Token (K)} $\downarrow$ & \textbf{Cost (\$)} $\downarrow$ \\
    \midrule
    \multirow{6}{*}{ExecutionAgent} 
        & Total   & 36.00 \textcolor{gray}{(18/50)} & \underline{27} & 422 & 0.131
                 & 18.00 \textcolor{gray}{(9/50)} & \textbf{31} & 479 & 0.078 \\
        & Python & \underline{54.55} \textcolor{gray}{(6/11)}   & \underline{29} & 461 & 0.140
                 & 27.27 \textcolor{gray}{(3/11)}   & \textbf{22} & 283 & 0.048 \\
        & C/C++  & 33.33 \textcolor{gray}{(6/18)}   & \underline{30} & 523 & 0.162
                 & 16.67 \textcolor{gray}{(3/18)}   & 44 & 739 & 0.120 \\
        & Java   & 44.44 \textcolor{gray}{(4/9)}    & \underline{21} & 427 & 0.131
                 & 11.11 \textcolor{gray}{(1/9)}    & \underline{28} & 423 & 0.069 \\
        & JS/TS  & 18.18 \textcolor{gray}{(2/11)}   & \textbf{22} & 219 & 0.071
                 & 18.18 \textcolor{gray}{(2/11)}   & \textbf{23} & 329 & 0.054 \\
        & Rust   & 0.00 \textcolor{gray}{(0/1)}     & 37 & 365 & 0.106
                 & 0.00 \textcolor{gray}{(0/1)}     & 35 & 370 & 0.059 \\
    \midrule
    \multirow{2}{*}{Repo2Run}
        & Total  & 54.54 \textcolor{gray}{(6/11)} & \textbf{15} & 847 & 0.238
                 & 27.27 \textcolor{gray}{(3/11)} & 43 & 1356 & 0.206 \\
        & Python & 54.54 \textcolor{gray}{(6/11)} & \textbf{15} & 847 & 0.238
                 & 27.27 \textcolor{gray}{(3/11)} & 43 & 1356 & 0.206 \\
    \midrule
    \multirow{6}{*}{Bash-based Agent}
        & Total   & -- & -- & -- & --
                 & \underline{36.00} \textcolor{gray}{(18/50)} & 33 & 466 & 0.072 \\
        & Python & -- & -- & -- & --
                 & 18.18 \textcolor{gray}{(2/11)}   & 44 & 794 & 0.122 \\
        & C/C++  & -- & -- & -- & --
                 & \underline{33.33} \textcolor{gray}{(6/18)}  & 36 & 467 & 0.072 \\
        & Java   & -- & -- & -- & --
                 & \textbf{66.67} \textcolor{gray}{(6/9)}  & \textbf{17} & \underline{155} & \underline{0.024} \\
        & JS/TS  & -- & -- & -- & --
                 & \underline{36.36} \textcolor{gray}{(4/11)}  & \underline{33} & 419 & 0.065 \\
        & Rust   & -- & -- & -- & --
                 & 0.00 \textcolor{gray}{(0/1)}  & 18 & 111 & 0.018 \\
    \midrule
    \multirow{6}{*}{RepoLaunch} 
        & Total   & \underline{56.00} \textcolor{gray}{(28/50)} & 29 & \textbf{185} & \textbf{0.054}
                 & 28.00 \textcolor{gray}{(14/50)} & 33 & \textbf{209} & \textbf{0.033} \\
        & Python & \underline{54.55} \textcolor{gray}{(6/11)}   & \underline{29} & \textbf{187} & \underline{0.054}
                 & 9.09 \textcolor{gray}{(1/11)}    & 41 & 278 & 0.044 \\
        & C/C++  & \underline{55.56} \textcolor{gray}{(10/18)}  & 31 & \underline{198} & \underline{0.057}
                 & 27.78 \textcolor{gray}{(5/18)}   & \underline{27} & \textbf{166} & \textbf{0.026} \\
        & Java   & \textbf{77.78} \textcolor{gray}{(7/9)}    & 22 & \underline{115} & \underline{0.033}
                 & \underline{33.33} \textcolor{gray}{(3/9)}    & 31 & 178 & 0.028 \\
        & JS/TS  & 45.45 \textcolor{gray}{(5/11)}   & 30 & \underline{203} & \underline{0.059}
                 & \textbf{45.45} \textcolor{gray}{(5/11)} & 36 & \underline{251} & \underline{0.039} \\
        & Rust   & 0.00 \textcolor{gray}{(0/1)}     & 50 & 380 & 0.109
                 & 0.00 \textcolor{gray}{(0/1)}     & \underline{12} & \textbf{51} & \textbf{0.008} \\
    \midrule
    \multirow{6}{*}{SWE-Agent}
        & Total   & \underline{56.00} \textcolor{gray}{(28/50)} & 29 & \textbf{185} & \textbf{0.054}
                  & 28.00 \textcolor{gray}{(14/50)} & 33 & \textbf{209} & \textbf{0.033} \\
        & Python  & 36.36 \textcolor{gray}{(4/11)}   & 38 & 266 & \textbf{0.042}
                  & 27.27 \textcolor{gray}{(3/11)}   & 45 & 366 & 0.057 \\
        & C/C++   & \underline{55.56} \textcolor{gray}{(10/18)}  & \underline{27} & \textbf{185} & \textbf{0.029}
                  & \underline{27.78} \textcolor{gray}{(5/18)}   & \underline{27} & \textbf{166} & \textbf{0.026} \\
        & Java    & 55.56 \textcolor{gray}{(5/9)}    & 32 & 253 & 0.040
                  & \underline{33.33} \textcolor{gray}{(3/9)}    & 31 & 178 & 0.028 \\
        & JS/TS   & 36.36 \textcolor{gray}{(4/11)}   & 36 & 251 & \textbf{0.039}
                  & \textbf{45.45} \textcolor{gray}{(5/11)} & 36 & \underline{251} & \underline{0.039} \\
        & Rust    & 0.00 \textcolor{gray}{(0/1)}     & \textbf{12} & \textbf{51} & \textbf{0.008}
                  & 0.00 \textcolor{gray}{(0/1)}     & \underline{12} & \textbf{51} & \textbf{0.008} \\
    \midrule
    \multirow{6}{*}{EnvPilot (Ours)} 
        & Total   & \textbf{72.00} \textcolor{gray}{(36/50)} & \underline{26} & \underline{198} & \underline{0.057} 
                  & \textbf{46.00} \textcolor{gray}{(23/50)} & \underline{32} & \underline{227} & \underline{0.036} \\
        & Python  & \textbf{63.64} \textcolor{gray}{(7/11)}   & 36 & \underline{278} & 0.080 
                  & \textbf{36.36} \textcolor{gray}{(4/11)}  & \underline{38} & \underline{266} & \underline{0.042} \\
        & C/C++   & \textbf{72.22} \textcolor{gray}{(13/18)}  & \textbf{26} & 195 & 0.056 
                  & \textbf{55.56} \textcolor{gray}{(10/18)} & \underline{27} & \underline{185} & \underline{0.029} \\
        & Java    & \textbf{77.78} \textcolor{gray}{(7/9)}   & \textbf{15} & \textbf{95} & \textbf{0.028} 
                  & \underline{55.56} \textcolor{gray}{(5/9)}   & 32 & 253 & 0.040 \\
        & JS/TS   & \textbf{72.73} \textcolor{gray}{(8/11)}   & \underline{28} & 206 & \underline{0.059} 
                  & \underline{36.36} \textcolor{gray}{(4/11)} & 36 & \underline{251} & \underline{0.039} \\
        & Rust    & \textbf{100.00} \textcolor{gray}{(1/1)}  & \underline{27} & \underline{235} & \underline{0.068} 
                  & 0.00 \textcolor{gray}{(0/1)}  & \textbf{12} & \textbf{51} & \textbf{0.008} \\
  \bottomrule
  \end{tabular}
  \end{adjustbox}
  \vspace{2pt}
  \begin{flushleft}
    \footnotesize \textbf{Note:} Bash-based Agent and Installamatic encountered technical issues with DeepSeek-V3-0324 as the model does not support the specific tool-calling framework. Repo2Run is evaluated only on the Python subset, as its specialized toolset is specifically designed for Python environments and lacks compatibility with other programming languages. The experimental results for DeepSeek-V3 and GPT-4o-mini are based on a 50-task subset (EnvBench Easy Subset).
  \end{flushleft}
\end{table}

\end{document}